\documentclass[superscriptaddress,groupedaddress,nofootnoteinbib,11pt]{article}
\pdfoutput=1 % if your are submitting a pdflatex (i.e. if you have images in pdf, png or jpg format)
\usepackage{jheppub}
\usepackage{graphicx,color,amsmath}
\usepackage{mathtools,slashed}
\usepackage{graphicx}% Include figure files
\usepackage[caption=false]{subfig}
\usepackage{dcolumn}% Align table columns on decimal point
\usepackage{multirow}
\usepackage{orcidlink}
\usepackage{tabularx}
\usepackage{bm}% bold math
\usepackage{hyperref}% add hypertext capabilities
\usepackage{amsmath}
\usepackage{amssymb}
\usepackage{comment}
\usepackage{setspace}

\usepackage{color}
\usepackage{enumerate}
\usepackage{booktabs}
\usepackage{siunitx}
 \graphicspath{{figures/}}

\newcommand{\nn}{\nonumber}
\newcommand{\E}{\mathcal{E}}
\newcommand{\lb}{\langle}
\newcommand{\rb}{\rangle_\beta}
\newcommand{\Lag}{\mathcal{L}}
\def\be{\begin{equation}}
\def\ee{\end{equation}}
\def\bea{\begin{eqnarray}}
\def\eea{\end{eqnarray}}

\newcommand{\muA}{\mu_{\rm applied}}
\newcommand{\muD}{\mu_{\rm dist}}
\newcommand{\muT}{\mu_{\rm total}}

\definecolor{mypurple}{RGB}{164,64,214}

\begin{document}

\title{Warm Inflation with Pseudo-scalar Couplings}

\date{\today}

%%%%%%%%%%%%%%%%%%%%%%%%%%%%%%%%%%%%%%%%%%%%%%%%%%%%%%%
\author{Edward Broadberry \orcidlink{0000-0003-0652-1862},}
\author{Anson Hook \orcidlink{0009-0005-7319-1439},}
\author{and Sagnik Mondal \orcidlink{0009-0007-4023-381X}}

\affiliation{Maryland Center for Fundamental Physics, Department of Physics, University of Maryland, College Park, MD 20742, U.S.A.}

\emailAdd{edbroad@umd.edu}
\emailAdd{hook@umd.edu}
\emailAdd{sagnik@umd.edu}

%%%%%%%%%%%%%%%%%%%%%%%%%%%%%%%%%%%%%%%%%%%%%%%%%%%%%%%

\abstract{Inflaton couplings during warm inflation result in the production of a thermal bath. Thermal friction and fluctuations can dominate the standard de Sitter analogues, resulting in a modified slow-roll scenario with a new source of density fluctuations. Due to issues with back-reaction, it is advantageous to consider inflaton couplings with the thermal bath that are pseudo-scalar in nature, e.g., derivative interactions or topological $F \tilde F$ couplings. We demonstrate that {\it every single} existing model of warm inflation utilizing pseudo-scalar couplings needs to be corrected to properly account for all of the chemical potentials that the thermal bath acquires in response to the inflaton coupling. These chemical potentials are for non-conserved charges, and are non-zero only because of the applied inflaton couplings.  The model-dependent chemical potentials modify the fluctuation-dissipation theorem, making the relationship between the thermal friction and thermal fluctuations model-dependent. In extreme cases, these chemical potentials can cause the friction term to vanish while thermal fluctuations remain non-zero. In the context of a simple example, we demonstrate how to calculate the chemical potentials, thermal friction, and thermal fluctuations using both the Boltzmann equations and by calculating thermal expectation values, showing explicitly that the two approaches give the same result.

}

\maketitle

\section{Introduction}

The observable universe has several intriguing features that are all elegantly solved by the hypothesis that the universe underwent an extended period of inflation~\cite{Guth:1980zm,Linde:1981mu}.  
At very large scales, the universe is approximately homogeneous with very low curvature.  
This homogeneity is extremely surprising when we realize that the many patches of space that we observe today appear to have never been in causal contact with one another.  
Requiring extremely homogeneous initial conditions is unappealing and begs a dynamical explanation.

Inflation solves all of the above problems by diluting away curvature and rapidly homogenizing the universe.  
An additional intriguing feature of inflation is that it can also explain the small $10^{-5}$ level in-homogeneities of the universe as well.  
Quantum fluctuations are blown up and expanded to cosmological scales and are the origin of all structure in the universe.  
The idea that our galaxy is only present due to a small quantum fluctuation of the early universe is an intriguing consequence of inflation.

Aside from all its virtues, inflation does have some flaws.  
The simplest and most appealing models of inflation are ruled out by measurements of the scalar-to-tensor ratio $r$~\cite{Planck:2018jri}.  
Inflation also tends to require field excursions that are $\mathcal{O}(1)$ super-Planckian, which may or may not be an issue from a more ultraviolet point of view.  
Finally, slow-roll inflation requires a potential that is somewhat flat.

Warm inflation provides an interesting take on cold inflation.  
The idea of warm inflation is that some coupling of the inflaton allows it to constantly lose energy populating a thermal bath~\cite{Berera:1995ie,Berera:1995wh,Berera:1996nv,Berera:1999ws} (for a review see~\cite{Berera:2008ar}).  %This energy loss generates a friction term that can dominate over Hubble friction.%
This energy loss generates a friction term which effectively adds to Hubble friction, even dominating over it in specific cases. 
Additionally, the thermal bath has classical thermal fluctuations, whose interactions with the inflaton can be the dominant source of its fluctuations.

On a more conceptual level, warm inflation provides a counterpoint where the observed density fluctuations are the consequence of classical thermal fluctuations, rather than quantum fluctuations.  
As such, it is a natural playground in which to explore experimental signals of the classical versus quantum nature of the origin of density perturbations.  
Additionally, the new source of friction can mitigate some of inflation's perceived flaws, such as the super-Planckian field excursions.

For all of its attractiveness, it has been notoriously difficult to write down a theory of warm-inflation.  
The original approach towards warm inflation model building involved utilizing scalar couplings between the inflaton and particles making up the thermal bath.  
Unfortunately, it was subsequently realized that for scalar couplings, friction terms are a sub-leading effect and the leading effect is to generate a thermal potential that makes inflation harder instead of easier~\cite{Yokoyama:1998ju}.

Over the years, there have been several different attempts to solve this problem.  
For example, one could endeavor to cancel the finite-temperature potential while keeping the friction term~\cite{Bastero-Gil:2016qru,Bastero-Gil:2019gao}.  
Another popular approach, the one we will focus on in this article, is the use of pseudo-scalar couplings to mitigate the generation of the thermal potential~\cite{Visinelli:2011jy,Mishra:2011vh,Ferreira:2017lnd,Ferreira:2017wlx,Kamali:2019ppi,Berghaus:2019whh}.  
These couplings are quite natural as there are other reasons to suspect that the inflaton might be a pseudo-scalar as opposed to a scalar, a scenario sometimes called natural inflation~\cite{Freese:1990rb}.

Once the pseudo-scalar couplings have been specified, the framework of linear response theory can be used to calculate the friction and/or the fluctuations~\cite{Berera:2002sp,Bastero-Gil:2010dgy}.  
Because the two are connected by the fluctuation dissipation theorem, usually only one of the two is calculated and the other is obtained from the theorem.  
Once the fluctuations are specified, the corresponding two-point function of the inflaton can be derived and compared to experiment.

In this article, we point out that pseudo-scalar interactions to leading order act like applied chemical potentials and, as such, the thermal bath acquires opposite signed chemical potentials to counterbalance the applied chemical potentials such that the total chemical potential is zero. 
The chemical potentials for non-conserved charges can drastically change both the friction and fluctuations, but have been neglected previously.  
All warm inflation models which contain pseudo-scalar interactions need to have their friction and fluctuations recalculated.

As an example of the drastic modification due to a pseudo-scalar interaction, consider the leading shift-symmetric interaction connecting an inflaton to any other sector :
\bea
\delta \mathcal{L} = \frac{\partial_\mu \phi}{f} J^\mu .
\eea
In the presence of a slow-rolling inflaton, $\dot \phi \ne 0$, this term modifies the Hamiltonian by  $\frac{\dot \phi}{f} Q$, which is clearly an applied chemical potential, we denote by $\muA$.  
The immediate response of the system is clear.  
It is energetically favorable to produce particles over anti-particles, and the thermal bath evolves towards a situation where there are more particles than anti-particles, which can be expressed in terms of a chemical potential $\muD$ that appears in the distribution function.  
In the process, it extracts energy from the inflaton, inducing friction.  
However, the end result of this process is also equally clear.  
In thermal equilibrium, the chemical potential for all non-conserved charges is zero.  
As such, the chemical potential present in the distribution functions, $\muD$, exactly balances out the applied chemical potential, $\muA$, so that in chemical equilibrium, the total chemical potential of the system, $\muT=\muA+\muD$, vanishes, implying $\muD=-\muA$. Once this occurs, the bath is in an equilibrium state and no more energy is extracted from the inflaton. 
The friction therefore vanishes up to Hubble-induced corrections.  
Fluctuations will remain non-zero as the inflaton is still coupled to the bath. 

What the previous example shows is that the backreaction of the thermal bath on the applied inflaton interaction can drastically change the resulting friction terms.  
The fact that the thermal bath will acquire chemical potentials for a generic pseudo-scalar interaction is to be expected from symmetries. 
Chemical potentials are $CPT$ odd.  
When a pseudo-scalar is expanded around $\dot \phi \ne 0$, this vacuum expectation value (vev) combined with $\phi$'s CP breaking interactions mean that $\dot \phi/f$ acts as a CPT-breaking spurion.  
Thus it is actually expected that unless other symmetries forbid it, this background will generate chemical potentials $\muD \sim \dot \phi/f$.

Finally, there is a useful analogy to be made between warm inflation and stirring water with a straw.  Water in a cup will evolve towards having no net velocity.  Now consider a person stirring the water with a straw.  Initially there will be friction on the straw.  However, experience shows that eventually the water will start co-rotating with the straw and the friction will drastically decrease and in some limit vanish.  This process does not change the thermal fluctuations present in the water.

In the context of warm inflation, a bath will naturally evolve towards having no net asymmetry between particles and anti-particles as long as number violating processes are present (water wants to be at rest).  Now consider the situation where the inflaton interacts via an applied chemical potential (stirring the water with a straw).  Initially there is friction that can be calculated with linear response theory, $\gamma_0$, and is the standard result used for friction in the literature (the initial friction on the straw).  However, eventually the bath responds with its own chemical potential, $\muD$, (water co-rotating with the straw) and the friction is suppressed and possibly vanishes, $\gamma = (1-d) \gamma_0$.  This entire process does not change the amplitude of thermal fluctuations, demonstrating that it is possible to get fluctuations that are not proportional to the equlibrium value of friction.

In Sec.~\ref{Sec: toy}, we work through a simple toy model to show how to calculate the chemical potentials developed by the bath in response to the inflaton interactions, as well as how to calculate the thermal fluctuations and the thermal friction on the inflaton.  
We do this both from the point of view of the Boltzmann equations, as well as simply by calculating a thermal expectation value of the relevant interaction in the equation of motion of $\phi$.  
Next, in Sec.~\ref{Sec: other}, we go through other pseudo-scalar interactions showing that each of them also generates chemical potentials that need to be taken into account properly.

Our results mean that the friction terms are smaller than previously calculated.  
As such, in Sec.~\ref{Sec: Example} we show that warm inflation can still work with a $\phi^4$ inflaton potential.  
Finally, we conclude in Sec.~\ref{Sec: conclusion}.

In our appendices, we review a great deal of background for those who have not done the relevant calculations before.  
In App.~\ref{App: thermal}, we review the thermal field double that we use to calculate thermal expectation values.  
We calculate expectation values using the in-in formalism that we review in App.~\ref{App: in-in}.  
Since the fluctuation dissipation theorem is often used to relate the thermal friction and fluctuation terms, in App.~\ref{App: fluc} we show that our results are consistent with this theorem once chemical potentials are included.  
In App.~\ref{App: 2pt}, we review the calculation of the inflaton 2-point function.  
In the context of warm inflation, previous calculations of the friction typically used linear response theory and the adiabatic approximation.  
In App.~\ref{App: linear}, we show how linear response theory is connected to how we calculate the friction.

\section{A Simple Toy Model} \label{Sec: toy}

In this section, we present a simple toy model of warm inflation within which we will perform all of our calculations.  
The model will be chosen to make the calculations as simple as possible.  
The backbone of warm inflation is an equation of motion of the inflation that obeys
\bea \label{Eq: slowrollstart}
\ddot \phi + (3 H + \gamma) \dot \phi + V'(\phi) = \xi.
\eea
There is a new source of friction, $\gamma$, which may or may not be larger than Hubble, and there is an extra term $\xi$ on the right hand side.  
The expectation value of $\langle \xi \rangle = 0$, so it does not affect slow roll.  
However, its two-point function is non-zero, $\langle \xi(x) \xi(y) \rangle \propto \delta^4(x-y)$, and can dominate over the vacuum fluctuations that typically determine the inflaton two-point function.  
Our goal in this section is to write a model from which these two features can be easily calculated.

The thermal bath from which friction and fluctuations originate consists of a single complex scalar, $\chi$, whose interactions are
\bea \label{Eq: V}
\mathcal{V}_{\rm int} = m_\chi^2 |\chi |^2 +  \lambda_\chi | \chi |^4 + \frac{\alpha_1}{4!} \chi^4 e^{4 i \phi/f} + \frac{\alpha_2}{4!} \chi^4 e^{5 i \phi/f} + h.c. 
\eea
Thermal equilibrium of the bath is maintained by the large coupling $\lambda_\chi \sim \mathcal{O}(1)$, while the couplings to the inflaton $\alpha_{1,2}$ are chosen to be small, and $m_\chi \ll T$ is chosen so that the thermal bath acts like radiation with $w = 1/3$.

We will additionally be taking the chemical potential $\muD \ll m_\chi$, so that the induced chemical potential does not cause the formation of a Bose-Einstein condensate.  
For simplicity, we will take $\alpha_{1,2}$ to be real. 
In what follows, we will be working to leading order in $\alpha_{1,2}$ and taking $\muD,\dot \phi/f \ll T$.

Note also that the inflaton coupling is chosen to be such that if either $\alpha_{1,2}$ are zero, then the inflaton is derivatively coupled $\grave{a}$ la a chemical potential. 
To see this, let's consider the case where $\alpha_2=0$.  In this situation, we can do the field redefinition $\chi \rightarrow e^{-i \phi/f} \chi$.  $\phi$ reappears from the kinetic term as $ \frac{\partial_\mu \phi}{f} J^\mu = \frac{\dot \phi}{f} J^0 \sim \muA Q$.  Expanding around the slow-roll background shows that in this limit, the $\alpha_1$ coupling acts like a chemical potential.

From the potential shown in Eq.~\ref{Eq: V}, we arrive at the equation of motion for the inflaton
\bea \label{Eq: phi eom}
\ddot \phi + 3 H \dot \phi + V'(\phi) = - 4 \frac{i}{f} \frac{\alpha_1}{4!} \chi^4 e^{4 i\phi/f} - 5 \frac{i}{f} \frac{\alpha_2}{4!} \chi^4 e^{5 i\phi/f} + h.c. \equiv O.
\eea
To make a connection with Eq.~\ref{Eq: slowrollstart}, it will be convenient to separate the RHS into its expectation value and fluctuation as
\bea
\xi = O - \langle O \rangle.
\eea
In this section, we will use the Boltzmann equation to calculate both the chemical potential, $\muD$, of the radiation bath, and the induced friction on the inflaton. 
We will then use thermal perturbation theory to calculate
\be\label{eq:FricDef}\langle O \rangle = - \gamma \dot \phi,\ee
and find agreement with the result obtained through the Boltzmann equation.

Next, we calculate the two-point function $\langle \xi(x) \xi(y)  \rangle$, which will ultimately determine the contribution of thermal fluctuations to the power spectrum. 
We will find that it agrees with the fluctuation-dissipation theorem, but only after chemical potentials have been included.  

%We will first use the Boltzmann equation to calculate the chemical potential $\mu$ of the thermal bath induced by this interaction.  In the process we will obtain the friction on the inflaton induced by the thermal bath.  Next we will switch gears and use thermal perturbation theory to calculate $\langle O \rangle = - \gamma \dot \phi$ to obtain the friction induced by the thermal bath.  Finally, we will again use thermal perturbation theory to calculate the two point function $\langle \xi(x) \xi(y)  \rangle$ to see the effects of the thermal bath on the inflaton.  Unsurprisingly the results of thermal perturbation theory will agree the results coming from the Boltzmann equation.

\subsection{Response of the Thermal Bath } 

%We first calculate how the thermal bath responds to the inflaton coupling in the limit where Hubble is small compared to all time scales in the problem.  
We first calculate how the thermal bath responds to the inflaton coupling in the limit where the Hubble time is large compared to all time scales in the problem.
As is well known, there are only a limited number of ways in which a thermal bath can respond to its external environment, such as by adjusting its temperature and chemical potential.  
In the simple toy model we consider, there is only a single complex scalar present, so that the only chemical potential possible is $\chi$'s chemical potential, $\muA$.

To find how $\muD$ and $T$ respond to a constant $\dot \phi$ background, we utilize the Boltzmann equations. 
First focus on the evolution of the number density due to $\alpha_1$. 
During slow-roll inflation, the inflaton coupling can inject or remove energy using
\begin{equation}
    \mathcal{L}_{\rm int} = -\frac{\alpha_1}{4!}e^{4i\frac{\dot{\phi}}{f}t}\chi^4,
\end{equation}
as the $e^{4i\frac{\dot{\phi}}{f} t}$ factor modifies the energy-conserving delta function in the Boltzmann equation. 
At leading order in $\dot{\phi}/Tf$ and $\muD/T$, we only need to consider two-to-two scattering.
\be
\chi(p_1)+\chi(p_2) \rightleftharpoons \chi^\dagger(p_3) + \chi^\dagger(p_4), 
\ee
for which the Boltzmann equation is
\bea
\left ( \frac{d \Delta n_\chi}{d t} \right )_{\alpha_1} 
&=& 4\alpha_1^2 \prod_{i,f}\int d \Pi_i d \Pi_f (2 \pi)^4 \delta^3(\vec{p}_{i}-\vec p_f) \delta\biggl(E_i-E_f+ 4 \frac{\dot \phi}{f}\biggr)\nn \\
&&\hspace{2cm}\times\bigl(\bar{f}_3 \bar{f}_4 (1+f_1) (1+f_2)   - f_1 f_2 (1+\bar{f}_3) (1+\bar{f}_4)  \bigr) , \label{Eq: ndot} %\\
%&\approx& 4 \int \prod_{i=1}^4 d \Pi_i (1+f_i) (2 \pi)^4 \delta^3(\sum_i \vec p_i) \delta(\sum_i E_i) \alpha_1^2 4 e^{-\beta (E_3 + E_4) } \left ( \frac{\dot \phi}{f} - \mu \right ) \nn \\
%&\equiv& 4 c_1 \alpha_1^2 \left ( 4 \frac{\dot \phi}{f} - 4 \mu \right ), \label{Eq: dndt}
\eea
where $\Delta n_\chi = n_\chi - n_{\chi^\dagger}$ and the distribution functions are given by
\bea
\label{eq:disfuncs1}
f_i &=& \frac{1}{e^{\beta(\omega_{i}-\muD)}-1}\implies \frac{1+f_i}{f_i}=e^{\beta(\omega_i-\muD)},\\
\bar{f}_i &=& \frac{1}{e^{\beta(\omega_i+\muD)}-1}\implies \frac{1+\bar{f}_i}{\bar{f}_i}=e^{\beta(\omega_i+\muD)}.
\label{eq:disfuncs2}
\eea
Because $\alpha_{1,2} \ll \lambda_\chi$, kinetic equilibrium is maintained faster than chemical equilibrium and the distribution functions are the standard ones allowed by the Boltzmann H theorem.
These identities allow us to expand to first order in $\dot{\phi}/Tf$ and $\muD/T$ :
\bea
\left ( \frac{d \Delta n_\chi}{d t} \right )_{\alpha_1}&\approx& 4\alpha_1^2\beta\left(4\frac{\dot{\phi}}{f}-4\muD\right) \int \prod_{i=1}^4 d \Pi_i f_1f_2\bar{f}_3\bar{f}_4 e^{\beta(\omega_3+\omega_4)}(2 \pi)^4 \delta^4(\sum_i p_i)\\
&\equiv& 4 c_1 \alpha_1^2 T^3\left ( 4 \frac{\dot \phi}{f} - 4 \muD \right ).\label{Eq: dndt}
\eea
The phase space integral, $c_1$, can be calculated numerically. 
When $m_\chi \lesssim T$, we find that $c_1 \approx (0.0589 -0.0632 \log (m/T)-0.143 e^{-m/T})/(2 \pi)^4 $ is a good numerical fit; namely, $c_1$ is an $\mathcal{O}(1)$ number divided by a phase space factor of $(2 \pi)^4$.
%When $m_\chi \lesssim 0.1 T$, we find $c_1 \approx (-0.526 -0.128 \log (m/T))/(2 \pi)^5 $ is a good numerical fit, namely $c_1$ is an $\mathcal{O}(1)$ number divided by a phase space factor of $(2 \pi)^5$.
A similar computation for $\alpha_2$ gives the total asymmetry of $\chi$ evolving as
\bea \label{Eq: chem pot dt}
\frac{d \Delta n_\chi}{d t} &=&  4 c_1 \alpha_1^2 T^3 \left ( 4 \frac{\dot \phi}{f} - 4 \muD \right ) + 4 c_1 \alpha_2^2 T^3 \left ( 5 \frac{\dot \phi}{f} - 4 \muD \right ) .
\eea
The evolution in time of the energy density is 
\bea \label{Eq: rho dt}
\frac{d \rho}{d t} =  4 \frac{\dot \phi}{f} c_1 \alpha_1^2 T^3 \left ( 4 \frac{\dot \phi}{f} - 4 \muD \right ) + 5 \frac{\dot \phi}{f} c_1 \alpha_2^2 T^3\left ( 5 \frac{\dot \phi}{f} - 4 \muD \right ) ,
\eea
which can be simply understood as the $\alpha_1$ ($\alpha_2$) processes depositing/removing $4 \dot \phi/f$ ($5 \dot \phi/f$) energy respectively.  
Because $\Delta n_\chi \sim \muD T^2$ and $\rho \sim T^4$, Eqs.~\ref{Eq: chem pot dt} and ~\ref{Eq: rho dt} imply that $\dot T/T \ll \dot{\mu}_{\rm dist}/\muD$.  
Thus, while the thermal bath will slowly heat up, $\dot T > 0$, it will quickly reach chemical equilibrium with $\dot{\mu}_{\rm dist} = 0$, giving
\bea
\muD &=& \frac{\alpha_1^2 + \frac{5}{4} \alpha_2^2}{\alpha_1^2 + \alpha_2^2} \frac{\dot \phi}{f},\\
\frac{d \rho}{d t} &=& \frac{\alpha_1^2 \alpha_2^2}{\alpha_1^2 + \alpha_2^2} c_1 T^3\left(\frac{\dot \phi}{f} \right)^2. \label{Eq: boltz fric}
\eea
In the limit that $\alpha_1$ or $\alpha_2$ are zero, the inflaton coupling is a chemical potential coupling, and we find the expected result that the system applies an equal and opposite dynamically generated chemical potential and the system reaches an equilibrium where $\dot T = \dot{\mu}_{\rm dist} = 0$.

Amusingly, at this point we have already calculated both the friction and fluctuations. 
Eq.~\ref{Eq: boltz fric} describes how energy is deposited into the radiation bath as a function of time.  
Conservation of energy dictates that this energy comes from the kinetic energy of the inflaton $\phi$, and thus we have already obtained the friction that the inflaton experiences.  
By the fluctuation dissipation theorem, discussed in more detail in App.~\ref{App: fluc}, this friction can be transformed into the inflaton fluctuations.

\subsection{Thermal Friction} \label{Sec: thermalfric}

In this subsection, we discuss how to calculate the friction term.  
In the previous subsection, we calculated the energy deposition in the process of calculating the chemical potential using the Boltzmann equation.  
However, it is always useful to calculate the same quantity in multiple ways.  
Thus, in this subsection, we directly calculate $\langle \mathcal{O}\rangle$ from Eq.~\ref{eq:FricDef} using thermal perturbation theory, reviewed in App.~\ref{App: thermal}.  
We start with the in-in correlator~\footnote{Since we are studying an equilibrium configuration, we could also use the in-out correlator and get the exact same result.  The in-in correlator happens to be easier to calculate.  See App.~\ref{App: in-in} for more details.}
\bea
\langle O \rangle =  \langle - 4 \frac{i}{f} \frac{\alpha_1}{4!} \chi^4 e^{4 i\phi/f} - 5 \frac{i}{f} \frac{\alpha_2}{4!} \chi^4 e^{5 i\phi/f} + h.c. \rangle.
\eea
Let us focus on the $\alpha_1$ term, as the $\alpha_2$ calculation will be analogous.  
To leading order in the interactions, we have
\bea
 \langle O_{\alpha_1} \rangle &=& - 4 \frac{i}{f} \frac{\alpha_1}{4!}  \langle i \int d^4 y \, [\chi(x)^4 e^{4 i \frac{\dot \phi}{f} t_x},\mathcal{L}_{\rm int}(y) ] \rangle + h.c.
\eea
Note that due to the explicit time dependence of the Lagrangian, many of the terms will vanish as one averages over times scales longer than $1/(\dot \phi/f)$.  
Meanwhile, other terms can be absorbed into the definition of the (thermal) potential of $\phi$.  
The only remaining term in this expectation value is the term coming from $\mathcal{L}_{\rm int}(y)$ whose time dependence is exactly $e^{-4 i \frac{\dot \phi}{f} t_y}$.  
Using this, we find that
\bea
\langle O_{\alpha_1} \rangle &=& - \frac{4}{f} \frac{\alpha_1^2}{4!}  \int d^3 y \int_{-\infty}^{t_x} d t_y e^{4 i \frac{\dot \phi}{f} (t_x - t_y)} \left (\langle \chi(x) \chi^{\dagger}(y) \rangle^4 - \langle \chi^{\dagger}(y) \chi(x) \rangle^4 \right )  + h.c. \nn \\
 &=& - \frac{4}{f} \frac{\alpha_1^2}{4!}  \int d^3 y \int_{-\infty}^{t_x} d t_y \int \prod_j \frac{d^3 k_j}{(2 \pi)^3 2\omega_i} e^{4 i \frac{\dot \phi}{f} (t_x - t_y)}\left\{ \left ( (1 + f_j) e^{-i k_j\cdot (x-y)}  + \bar{f}_je^{i k_j\cdot (x-y)} \right )^4\right. \nn\\&&\hspace{4cm} \left. - \left ( (1 + \bar{f}_j) e^{i k_j\cdot(x-y)}  +  f_j e^{-i k_j\cdot (x-y)}  \right )^4 \right\}  + h.c. \label{Eq: 1 point},
\eea
where in the second line we have used the thermal field propagators in Eq.~\ref{eq:Prop1} and  Eq.~\ref{eq:Prop2}.

There are several facts that can be used to simplify this expression.  
Firstly, the $d^3 y$ integral just gives momentum-conserving delta functions.  
While not as obvious, something similar happens for the time integral over $t_y$.  
$t_y$ is only integrated over the half plane so that instead of giving a delta function in energy, its imaginary part is a principle value while its real part is half the typical delta function.  
The effect of adding the hermitian conjugate is to remove the principle value piece and just give the energy-conserving delta functions modified to include the energy deposition of the inflaton.

The rest of Eq.~\ref{Eq: 1 point} has interpretations coming from different terms in the Boltzmann equations.
For example, the piece containing $(1 + f_j)^4$ describes the process $0 \rightarrow \chi^4$, a process energetically allowed only at higher orders in $\dot \phi/f$, but is nevertheless still present. 
At leading order, however, we obtain the Boltzmann equation for two-to-two scattering from the previous subsection. 
Putting it all together, including the symmetry factors for the phase space of identical particles, we find
\bea
\langle O \rangle &=& - \frac{4 c_1}{f} \alpha_1^2T^3 \left ( 4 \frac{\dot \phi}{f} - 4 \muD \right ) - \frac{5 c_1}{f} \alpha_2^2T^3 \left ( 5 \frac{\dot \phi}{f} - 4 \muD \right ) = - \frac{\alpha_1^2 \alpha_2^2}{\alpha_1^2 + \alpha_2^2}c_1 \frac{T^3}{f^2} \dot \phi \nn, \\
\gamma &=& c_1\frac{\alpha_1^2 \alpha_2^2}{\alpha_1^2 + \alpha_2^2} \frac{T^3}{f^2} . \label{Eq: gamma}
\eea
In our model, $\alpha_{1,2}$ are dimensionless so that $\gamma \sim \alpha_{1,2}^2 T^3$ by dimensional analysis.  If instead $\chi^3$ ($\chi^2$) interactions were being considered, then again dimensional analysis would give $\gamma \sim \alpha_{1,2}^2 T$ ($\gamma \sim \alpha_{1,2}^2/T$).  This shows that simple modifications of our model will be able to accommodate different temperature scalings of the friction.

A good double check of our results is that
\bea
\langle O \rangle &=& - \frac{d\rho/dt}{\dot \phi}
\eea
holds even when the chemical potential is not at its equilibrium value and without needing to take the $\muD, \dot \phi/f \ll T$ limit.  It is refreshing that while our friction coefficient was calculated using thermal perturbation theory, and energy deposition was calculated using the Boltzmann equations, the results agree.

\subsection{Two Point Function}

We now use thermal perturbation theory to calculate the two point function $\langle \xi(x) \xi(y)  \rangle$.  
The first thing to note is that it involves a propagator from $x$ to $y$.  
Famously, position space propagators will fall off exponentially with correlation length and time.  
As such, as long as we are interested in distances much larger than the correlation length, it is reasonable to approximate
\bea
\langle \xi(x) \xi(y)  \rangle \approx A \delta^4(x-y) .
\eea
In the rest of the subsection, we will calculate the constant $A$. Selecting out the pieces that do not oscillate quickly and average away, we have
\bea
A &=& \int d^4 y \langle \xi(x) \xi(y)  \rangle \\
&=& \int d^4 y \frac{1}{4! f^2} \big ( \langle \chi(x) \chi^\dagger(y) \rangle^4 \left (16 \alpha_1^2 e^{4 i \frac{\dot \phi}{f} (t_x - t_y)} + 25 \alpha_2^2 e^{5 i \frac{\dot \phi}{f} (t_x - t_y)} \right) \nn \\
&+& \langle \chi^\dagger(x) \chi(y)  \rangle^4 \left (16 \alpha_1^2 e^{-4 i \frac{\dot \phi}{f} (t_x - t_y)} + 25 \alpha_2^2 e^{-5 i \frac{\dot \phi}{f} (t_x - t_y)} \right) \big ). \label{Eq: A}
\eea
The only difference between this equation and our previous Eq.~\ref{Eq: 1 point} comes down to a minus sign. 
Due to the forward and backward process depositing or removing energy, Eq.~\ref{Eq: 1 point} has a piece $e^{- \beta (4 \dot \phi/f - 2\muD)}- e^{-\beta 2 \muD} \approx \beta (4 \muD - 4 \dot \phi/f)$. 
Due to factors of $i$ difference, the current two point function, Eq.~\ref{Eq: A}, has $e^{- \beta (4 \dot \phi/f - 2\muD)}+ e^{-\beta 2 \muD} \approx 2$. 

As a result, after a bit of algebra we find that
\bea \label{Eq: fluc}
\langle \xi(x) \xi(y)  \rangle \approx \frac{2 c_1 T^4}{f^2} \left ( 16 \alpha_1^2 + 25 \alpha_2^2 \right ) \delta^4(x-y);  \qquad A = \frac{2 c_1 T^4}{f^2} \left ( 16 \alpha_1^2 + 25 \alpha_2^2 \right ).
\eea
We can combine this result with the friction term from the previous subsection to find
\bea
A &=&2\left(\gamma  \frac{16\alpha_1^2+25\alpha_2^2}{16\alpha_1^2\left(1-\frac{\muD}{\dot{\phi}/f}\right)+25\alpha_2^2 \left(1-\frac{4\muD}{5\dot{\phi}/f}\right)}\right)T\nonumber \\
&=& 2\gamma T \frac{(\alpha_1^2+\alpha_2^2)(16\alpha_1^2+25\alpha_2^2)}{\alpha_1^2\alpha_2^2}\label{eq:FlucFinal}.
\eea
While not immediately obvious, as discussed in App.~\ref{App: fluc}, this result for the fluctuation is what we expect from the fluctuation dissipation theorem. 
As can also be seen from Eq.~\ref{eq:FlucFinal}, if one were to neglect the chemical potential, $\muD = 0$, then the standard result of $A = 2 \gamma T$ would be obtained.
From our results, we see that zero chemical potentials is an important implicit assumption when it was originally claimed that the fluctuation dissipation theorem implies $A=2\gamma T$ \cite{Graham:2009bf,Bastero-Gil:2014raa}.

We have seen in this section that the chemical potential will always act to reduce the friction coefficient, which fits with the analogy of stirring a water cup, where the water develops a current that travels along with the stirrer. We also found that at leading order the chemical potential has no effect on the size of the fluctuations, which is also consistent with this analogy. It is therefore clear that in Eq.~\ref{eq:FlucFinal}, the fluctuation-dissipation theorem is modified not by enhancing the fluctuations, but by reducing the dissipation. This is clearly a general statement that is applicable to any model of warm inflation with a chemical potential.

\section{Other Examples of Chemical Potentials} \label{Sec: other}

In this section, we discuss examples other than our toy model in order to demonstrate the ubiquity of chemical potentials.  Because of the number of models being discussed, we will consider each example only qualitatively, though one can obtain the $\mathcal{O}(1)$ numbers in a manner similar to the previous section. As before, our analyses will be done in the $\dot \phi/f \ll T$ and $H \approx 0$ limit.

In the language of the water stirring analogy in the introduction, we find three main take-aways from these examples:
\begin{enumerate}
    \item The friction in chemical equilibrium is $\gamma$.  Neglecting chemical potentials, would lead one to calculate $\gamma_0$ and miss the factor of $1-d$.
    \item If one calculates fluctuations and uses the fluctuation-dissipation theorem to find the friction, one has to account for chemical potentials or else one will find $\gamma_0$ instead of $\gamma$.
    \item Finally, axion-like couplings lead to both perturbative and non-perturbative contributions to the friction.  %These frictional effects have the same scaling and the coefficients of both sources of friction need to be determined.
\end{enumerate}

While many of the couplings below will have $\gamma \approx 0$ due to a chemical potential when $\dot \phi/f \ll T$, this does not mean that it is impossible to do warm inflation with these couplings.  Some of the couplings have equilibrium configuration where $\dot \phi/f \sim \muD \sim T$.  Taking into account a non-zero Hubble, these couplings can support $\gamma \propto H$, possibly supporting warm inflation in the weak ($\gamma < H$) regime.

\subsection{Chemical Potential Couplings} \label{Sec: chemical}

The first example we will illustrate is a chemical potential coupling of a different form.  In particular, we will take the inflaton to be coupled to fermions $\psi_{1,2}$ with 
\bea
\mathcal{L}_{\rm int} = \frac{\partial_\mu \phi}{f} (J^\mu_{\psi_{1}} + J^\mu_{\psi_{2}}) ,
\eea
which amounts to a chemical potential favoring particles over anti-particles.  Additionally, we include a symmetry-breaking coupling of the form $\delta \mathcal{L} = \alpha \psi_1^2 \psi_2^2$.

There are two ways in which we can use this coupling.  Because this coupling is bilinear in the fermion fields, we can include it directly when decomposing the field in terms of creation and annihilation operators:
\bea
\psi_{1,2}(x) &=& \int \frac{d^3 p}{(2 \pi)^3 \sqrt{2 \omega_p}} \sum_s \left (u_s(p) a_{p} e^{-i p x - i \muD t} + v_s(p) b^\dagger_{p} e^{i p x - i \muD t} \right ).
\eea
This amounts to just changing the energy of particles and anti-particles to $E = \sqrt{p^2+m^2} \pm \dot \phi/f$.  
Note that in Eq.~\ref{Eq: ndot}, the effect of the coupling is purely to modify the energy conservation delta function, so it is clear that the results are the same whether one works in this basis or the other basis. 
Symmetry-breaking operators will cause a deposition of energy, and the induced thermal chemical potential will be such to exactly cancel the $\dot \phi/f$-induced energy differences.

A different approach to this coupling uses perturbation theory.  
Like when doing perturbation theory with a mass term, the previous answer can be obtained by doing a resummation of diagrams.  When doing this calculation, one needs to go to higher order in perturbation theory to see a non-trivial answer.  Schematically,
\bea
 _{\rm in} \langle \dot J^0_\psi(0) \rangle_{\rm in,\beta} \sim \lb \dot J^0_\psi(0) \frac{\dot \phi}{f} J_\psi^0(x) \rb + \alpha^2 \frac{\dot \phi}{f} \lb \dot J_\psi^0(0) J_\psi^0(x) \psi_1^2(y) \psi_2^2(y) \psi_1^2(z) \psi_2^2(z) \rb + \cdots
\eea
The effect of $\frac{\dot \phi}{f} J^0_\psi$ is to include the energy change $\pm \dot \phi/f$ for particles versus anti-particles at leading order in perturbation theory.  
However, notice that the leading term is zero. 
One needs interactions to create a connected diagram. 
This is equivalent to the statement that if $\alpha \rightarrow 0$, the chemical potential can be removed by a field redefinition.
In particular, one needs three different insertions, $\frac{\dot \phi}{f} J^0_\psi$ and two $\alpha \psi_1^2 \psi_2^2$ insertions, until a non-zero result is obtained.  
The expression obtained will be the fermionic analogue of Eq.~\ref{Eq: dndt}.

\subsection{Symmetry Breaking Couplings}

In this subsection, we will consider the interaction between the inflaton and scalar field, $\chi$, to be
\bea \label{Eq: symbreak}
\mathcal{V}_{\rm int} = \frac{\alpha_1}{4!} \chi^4 (1 + 4 i \frac{\phi}{f}) + h.c. 
\eea
Note that this interaction is similar to the previous Eq.~\ref{Eq: V} with $\alpha_2=0$, but only the leading order term in $1/f$ is included.  Because of the lack of higher-order terms, this is not a shift-symmetric chemical potential term.  However, it will still induce a chemical potential response from the bath.  
The reason is that when we did all of our calculations in Sec.~\ref{Sec: toy}, we eventually had to Taylor series our result in $\dot \phi/(f T)$. 
Because of this, the entirety of our calculation is completely insensitive to the exact form of the coupling $e^{4 i \phi/f}$ and only to the first-order term in the Taylor series.

The most important aspect of the coupling in Eq.~\ref{Eq: symbreak} is the fact that the symmetry-breaking term is a pseudo-scalar interaction and has a factor of $i$.  This $i$ is critical in making the particle versus anti-particle interactions different by a minus sign and thus gives rise to a chemical potential.

\subsection{Topological Axion Couplings}

The final couplings that we discuss are axion couplings to abelian ($F$) and non-abelian ($G$) gauge groups of the form
\bea  \label{Eq: axion}
\mathcal{V}_{\rm int} = \frac{\alpha \phi}{4f} F \tilde F + \frac{\alpha \phi}{4f} G \tilde G.
\eea
Both of these couplings have been used for warm inflation model building~\cite{Ferreira:2017lnd,Ferreira:2017wlx,Kamali:2019ppi,Berghaus:2019whh,Berghaus:2024zfg,Berghaus:2025dqi,Montefalcone:2022jfw}.
We will discuss the case where the gauge bosons are abelian or non-abelian separately.  As mentioned before, we will be qualitative about $\mathcal{O}(1)$ numbers, but we will keep track of $\alpha$ as $\alpha$ counting can help clarify issues when light fermions become involved.

\subsubsection{Abelian Gauge Groups - No Fermions:}

As is well known, the axion coupling changes the dispersion relation of the gauge boson to
\bea \label{Eq: dispersion}
E^2 = p^2 \pm \alpha \,p \frac{\dot \phi}{f} + m^2 ,
\eea
where we have allowed for a mass term to stabilize the tachyon if need be. 
The $\pm$ depends on if the gauge boson is left- or right-circularly polarized.  
In the absence of a thermal bath, one of the two polarizations can have a tachyonic mode that grows exponentially quickly.   The time scale of the exponential growth is $t \sim f/\dot \phi$.  This removal of energy is exponential in nature and thus too fast to be accounted for with a simple friction term $\gamma \dot \phi$.  Even without a thermal bath, it has been proposed to use this coupling to support inflation~\cite{Anber:2009ua,Anber:2012du}.

As it will be useful later on, observe that the dispersion relation in the limit of $p \sim T \gg \muD, m$ is
\bea \label{Eq: dis expand}
E = p \pm \alpha\frac{\dot \phi}{2 f} + \cdots .
\eea
This modification looks similar to a chemical potential, but also leads to a plethora of effects, such as polarization rotation, that are often used to look for axion dark matter.

\subsubsection{Abelian Gauge Groups - Fermions:}

An abelian gauge theory by itself does not thermalize~\footnote{If the inflaton comes into equilibrium, then thermal equilibrium can be obtained without fermions.  This situation was numerically studied in Ref.~\cite{Ferreira:2017lnd} where it was found that the equilibrium configuration contained a chemical potential for the helicity of the gauge boson.}, so let us consider what occurs when there is a light fermion, $m_e \ll T$, that we will lazily call the electron.  
In the presence of a thermal bath, the tachyonic instability can be cured, as a thermally-induced photon mass $m \sim e T$ will be generated that eventually removes the instability.
To obtain the friction, it is easiest to consider the Boltzmann Equation. 
As can be clearly seen from Eq.~\ref{Eq: dis expand}, the leading-order effect of the axion coupling is to apply an external chemical potential for helicity, much like the situation in Sec.~\ref{Sec: chemical}.
\begin{figure}[t]
\centering
\includegraphics[width=.4\linewidth]{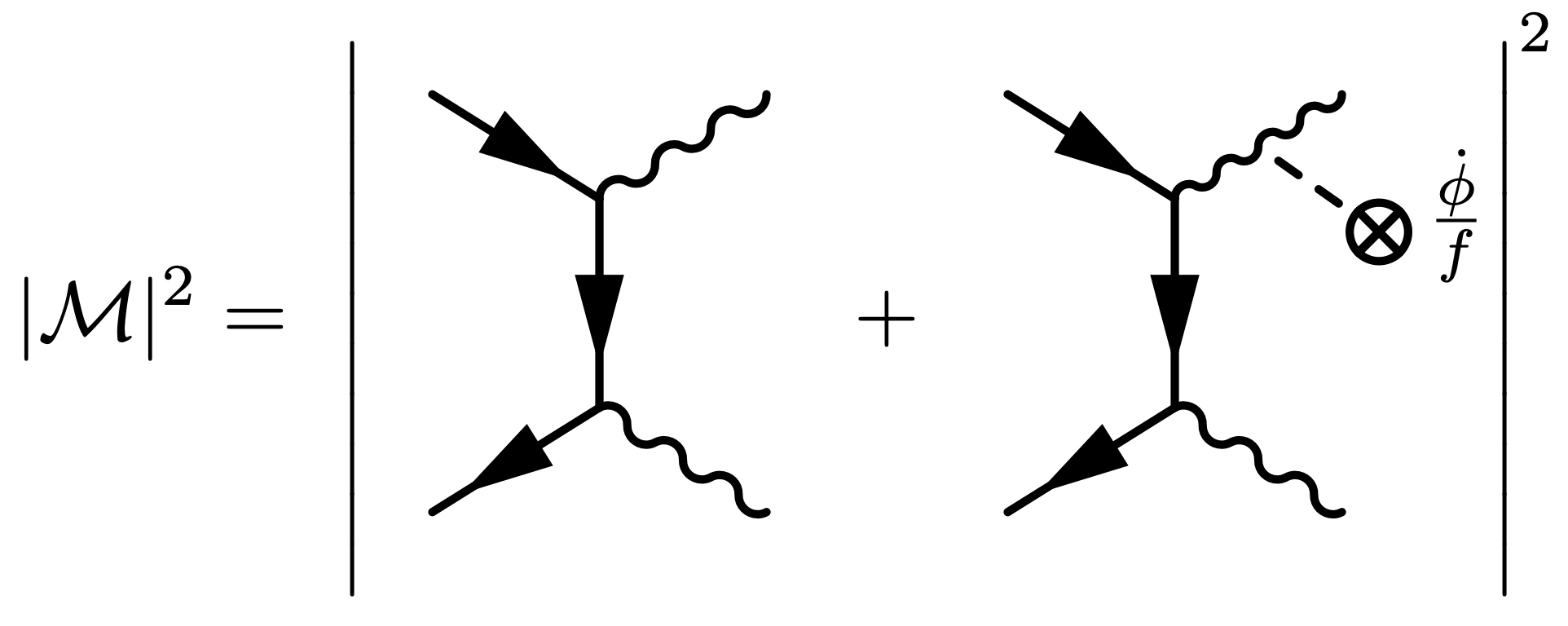}
\caption{The Feynman diagrams for the scattering process $e+e^- \rightarrow A_{L,R} + A_{L,R}$. The leading contribution in $\dot{\phi}/f$ comes from the cross term.}
\label{Fig: Nonab}
\end{figure}
To see that the system will develop a chemical potential in response, let us consider a scattering process $e^+ + e^- \rightarrow A_L + A_L$, writing the left(right)-handed gauge boson as $A_L$($A_R$), with Feynman diagrams shown in Fig.~\ref{Fig: Nonab}. 
This helicity-violating process requires a mass insertion.  
If one wishes to avoid the possible mass suppression, consider instead the process $e^+ + e^- \rightarrow 3 A$.  
The combination of these different processes makes it clear that the chemical potential that the system develops in response is the expected chemical potential for helicity of the gauge boson, and not a chemical potential for the helicity of the electron.

%The presence of a mass insertion complicates matters as this process also changes the helicity of the electron.  From this process alone, it is unclear if the theory develops a chemical potential for electron helicity, photon helicity or a combination of the two.  To break the degeneracy, consider the photon number changing process $A_R + e \rightarrow A_L + A_L + e$.  This process has an energy change of $3 \alpha \dot \phi/f$ but since the helicity change of the electron is no different than the case of the compton scattering, the solution to the chemical potential equations indicate that {\it no} chemical potential for the electron is developed and {\it only} a chemical potential for photon helicity is obtained.

The end result of these scattering processes is that
\bea \label{Eq: dardt}
\frac{d n_{A_R}}{dt} - \frac{d n_{A_L}}{dt} \sim (\alpha^2 m_e^2 T + \alpha^3 T^3) \times \left ( \frac{\alpha \dot \phi}{f} - \muD \right ) ,
\eea
where the factors of $T$ are found by dimensional analysis. 
The thermal bath will develop its own chemical potential $\muD \sim \alpha\dot \phi/f$ to counteract the applied chemical potential. 

As the system approaches chemical equilibrium, there will be a friction term $$\gamma \sim \alpha^4 m_e^2 T/f^2 \textrm{, or } \gamma \sim \alpha^5 T^3/f^2.$$  
However, once the system reaches chemical equilibrium, the friction will vanish, $\gamma \rightarrow 0$. 

The only energy deposition will now come from the fact that at higher orders in $\dot \phi/fT$, the coupling is not exactly a chemical potential.  
The appearance of more powers of $\dot \phi$ indicates that while there will be energy deposition, it will not take the simple form of a friction term.  
Interestingly enough, the thermal bath responds to the applied chemical potential even if the electron is massless.

It is useful to study what occurs when one performs a field redefinition $\psi \rightarrow e^{i \phi/f} \psi$ to move the topological coupling in Eq.~\ref{Eq: axion} into
\bea \label{Eq: twofermion}
\mathcal{L} = m e^{i \phi/f} \psi \psi^c - \frac{\partial_\mu \phi}{f} J^\mu .
\eea
In this basis, it appears as if everything is about the chirality of the electron rather than the chirality of the photon.  We will now describe how to show that, even in this basis, everything is about the chirality of the photon and not the electron.

In this basis, there are two effects.  Any fermion chirality violation necessarily involves both a mass insertion, which deposits energy $\dot \phi/f$ from the mass term, and at the same time removes an energy $\dot \phi/f$ from the derivative interaction.  Due to this minus sign, any process that changes the chirality of the fermion has no energy injection, and no chemical potential is obtained for the fermion helicity.  The two effects in Eq.~\ref{Eq: twofermion} cancel each other out exactly.  

The only effect of the combined couplings is on photon propagation via a loop of electrons.  
For example, the typical Compton scattering diagrams can interfere with analogous diagrams where one of the photons is dressed by electron loops, changing its dispersion relation.  This process would give the rate Eq.~\ref{Eq: dardt}.  As before, even if the electron is massless, a chemical potential for photon helicity still develops, shutting off the friction up to effects due to non-zero Hubble.

\subsubsection{Non-Abelian Gauge Groups - No Light Fermions:}

When discussing non-abelian gauge groups, there are both perturbative effects and non-perturbative effects.  
Let us discuss these contributions separately.

\subsubsection*{Perturbative Effects:} As with the abelian case, to leading order in the gauge coupling, $g$, the effect of Eq.~\ref{Eq: axion} is a chemical potential for helicity.  
At higher order in $g$, there are other effects.
As we are working perturbatively, the main difference between non-abelian and abelian gauge groups is that non-abelian groups contain interactions that automatically violate helicity conservation as well as thermalize the system.  
As such, we will discuss thermal equilibrium and the Boltzmann equation from the beginning.

Let us first discuss the effect of the applied chemical potential for helicity.  
The leading-order violation of helicity in pure Yang-Mills theory is from two-to-three scattering process such as $A_R + A_R \rightarrow A_R + A_R + A_R$.  
This leads to a rate
\bea \label{Eq: pureYM}
\frac{d n_{A_R}}{dt} - \frac{d n_{A_L}}{dt} \propto \alpha^3 T^4 \times \left  ( \frac{\alpha \dot \phi}{f T} - \frac{\muD}{T} \right ) .
\eea

Aside from a chemical potential, there are also $A^3 \dot \phi/f$ and $A^4 \dot \phi/f$ couplings.  These also contribute to helicity-violating processes at the same order as Eq.~\ref{Eq: pureYM}.  
Due to the different kinematic structure of the vertices caused by the $\epsilon$ tensor, the interference that generates Eq.~\ref{Eq: pureYM} is a bit more subtle, but the overall expectation is something of nearly the same order.

Unlike the abelian case, even at leading order in $\dot \phi/f$, the coupling is not just an applied chemical potential.  
The situation is similar to our toy  model, and we expect that after the system develops a chemical potential $\muD \propto \alpha \dot \phi/f$, there are still heating effects proportional to $\dot \phi/f$.  Namely, we expect that 
\bea \label{Eq: pureYM-bath}
\frac{d \rho_{\rm bath}}{dt} \propto \alpha^3 T^4 \times \frac{\alpha \dot \phi}{f T} \times \frac{\alpha \dot \phi}{f},
\eea
where the first term is the scattering rate, the second term is the bias, and the last term is how much energy is injected into the bath per event.  However, additional work is needed to make certain of this result.

\subsubsection*{Non-Perturbative Effects:}
Sphalerons are a non-perturative effect that has similar scaling to the perturbative effects considered above, but whose order-one numbers make it less suppressed.  The non-perturbative friction, assuming no chemical potentials that result from an axion coupled to a non-abelian gauge group, has been studied before in Ref.~\cite{McLerran:1990de,Laine:2016hma} giving the result $\gamma \sim \alpha^5 T^3/f^2$.  To better understand the origin of this friction and how a chemical potential for helicity can change it, we provide a simple understanding of the effect below.

Sphalerons are a process by which the Chern-Simons number changes by one with a rate $\sim \alpha^5 T^4$.  
Roughly speaking, sphalerons are CP-violating collective excitations of $1/\alpha$ particles.
%When dealing with particles, to lowest order in $\alpha$ the Chern-Simons number measures how many more $A_L$ there are than $A_R$ (up to a factor of order $ 1/\alpha$).  
Thus, to leading order, we can interpret sphalerons as objects that create $1/\alpha$ more left-handed than right-handed gauge bosons.  
As discussed before, the effect of the $\phi G \tilde G$ coupling is to introduce a chemical potential for helicity.  
As with Eq.~\ref{Eq: pureYM}, we find that the leading-order effect of sphalerons is
\bea \label{Eq: pureYM-NA}
\frac{d n_{A_R}}{dt} - \frac{d n_{A_L}}{dt} \propto \alpha^5 T^4 \times \frac{1}{\alpha} \left ( \frac{\alpha \dot \phi}{f T} - \frac{\muD}{T} \right ).
\eea
The rate is suppressed compared to Eq.~\ref{Eq: pureYM}; however, there is an enhancement that comes with each event, since it changes the helicity of $1/\alpha$ gauge bosons.

Before the chemical potential is taken into account, the energy deposited into the thermal bath mediated by sphalerons is
\bea
\frac{d \rho_{\rm bath}}{dt} \propto \alpha^5 T^4 \times \frac{1}{\alpha} \frac{\alpha \dot \phi}{f T} \times \frac{1}{\alpha} \frac{\alpha \dot \phi}{f},
\eea
giving the friction term found in Ref.~\cite{Laine:2016hma}.
Note that this is parametrically similar to the perturbative result in Eq.~\ref{Eq: pureYM-bath}, but it comes about in a different way.  The rate is suppressed, but since it deals with $1/\alpha$ particles, the end result is comparable.

The next step is to properly account for the effects of the chemical potential.  If we only consider sphalerons, then it is possible that the chemical potential will cancel all helicity violation (up to terms higher order in $\alpha$) and that friction will be suppressed.  It is also equally possible that, much like the perturbative result, the chemical potential will not cancel out the effect entirely.  Either way, eventually terms that are higher order in $\alpha$ or the perturbative effects will render the friction non-zero.  However, these effects are either suppressed by $\alpha$ or by many phase-space factors.
A dedicated study would be needed to determine the equilibrium configuration.

%The friction resulting from this is at the same order in $\alpha$ as the perturbative effects, so it must be corrected to include the effects of a chemical potential, should be subleading to the effects estimated above.

\subsubsection{Non-Abelian Gauge Groups - Light Fermions:}

\subsubsection*{Perturbative Effects :}
The presence of light fermions allow for new helicity-violating processes similar to the Abelian case.  As discussed there, Compton-type scattering will give a contribution to Eq.~\ref{Eq: pureYM} of order $\alpha^2 m_f^2 T \alpha \dot \phi/f$, while two-to-three scattering processes will give rise to terms that scale like Eq.~\ref{Eq: pureYM} up to Casimirs.

As before, these results can be understood in any basis.  In the basis with a $\phi G \tilde G$ coupling, the fermion scattering allows for gauge-boson helicity violation which extracts energy from the chemical potential portion of the coupling.  Additionally, $A^3$ and $A^4$ terms will contribute in a way that is not chemical potential like.  

In the basis where the axion coupling has been rotated into the light fermion couplings, a similar feature to what occurs in the abelian case abounds.  The mass and derivative interactions cancel each other out for almost all interactions except for processes where there is a loop of fermions that modify either the propagator, the three-point function, or the four-point function of the non-abelian gauge boson.

Regardless of basis, a chemical potential for the gauge boson helicity is formed and will drastically change the results.  As in the case of no fermions, a dedicated analysis will be required.

%On the approach towards equilibrium, any helicity violating process will induce energy deposition giving frictional terms proportional to these perturbative helicity violating rates.  These perturbative contributions to friction will dominate the previously considered non-perturbative contributions.  Because the axion coupling only at leading order is chemical potential like, whether or not the final equilibrium has significant friction will require a more detailed analysis.  

\subsubsection*{Non-Perturbative Effects :}

Non-perturbative frictional effects including fermions were first studied in Ref.~\cite{McLerran:1990de}.
Because of the anomaly, sphalerons are tied together with fermion helicity violation.  This effect is similar to how helicity changes of the photon and fermion are connected in Compton scattering.  As such, it is not at all clear what the equilibrium configuration is, as chemical potentials for both gauge boson and fermion helicity will need to be considered. 
Non-abelian warm inflation with light fermions has also been studied in Ref.~\cite{Berghaus:2020ekh,Drewes:2023khq,Berghaus:2024zfg,Berghaus:2025dqi}, where they included the chemical potential for fermions but not the gauge bosons and neglected the effect of the chemical potential on the fluctuation dissipation theorem when calculating the power spectrum. 
A dedicated calculation would be needed to elucidate what actually occurs in this situation.

\section{A warm $\phi^4$ example} \label{Sec: Example}

In the previous sections, we have discussed how friction and fluctuations are drastically changed by the presence of chemical potentials.  In this section, we demonstrate that warm inflation can still occur for a simple inflaton potential.

We first give a brief review of warm inflation before summarizing the results of a power spectrum calculation.  Next, we discuss the thermal potential, as our example model was chosen for calculational simplicity as opposed to suppressing the thermal potential.  Finally, we present a data point demonstrating that warm $\phi^4$ inflation is viable.

\subsection{The Background Solution}

The basic idea of warm inflation is to modify the usual slow-roll inflationary dynamics by coupling the inflaton to a radiation bath that's in thermal equilibrium. The background equations of motion are
\bea
    &&\ddot{\phi} + (3H+\gamma)\dot{\phi}+V'(\phi) = 0,\\
    &&H^2 = \frac{1}{3 M_{\rm pl}^2}\left(V(\phi) + \frac{1}{2}\dot{\phi}^2 + \rho_r\right),\\
    &&\dot{\rho}_r + 4H\rho_r = \gamma \dot{\phi}^2,
\eea
where the friction term, $\gamma$, is the method by which energy is constantly pumped from the inflaton into the radiation sector. The radiation energy density is $\rho_r = g_*\pi^2T^4/30 $ with $g_*$ being the number of degrees of freedom. Typically, one defines the ratio of the friction term to the Hubble parameter
\be Q \equiv \frac{\gamma}{3H}.\ee

The usual slow-roll parameters are modified to give
\bea
\epsilon = \frac{1}{1+Q}\frac{M_{pl}^2}{2} \left ( \frac{V'}{V} \right )^2, \qquad \eta = \frac{1}{1+Q}\frac{M_{pl}^2 V''}{V}.
\eea
If $\epsilon, |\eta|\ll 1$, then the approximate slow-roll background solutions are 
\bea
&&\dot{\phi}\approx - \frac{V'(\phi)}{3H+\gamma},\label{eq:sr1}\\
&&H^2 \approx \frac{V}{3 M_{\rm pl}^2},\label{eq:sr2}\\
&&\rho_r \approx \frac{\gamma}{4H} \dot{\phi}^2.\label{eq:sr3}
\eea
The third equation is the condition that the energy extracted from the inflaton per unit Hubble time is sufficient to keep the radiation sector in thermal equilibrium. 

In the rest of this section, we show that within our simple model it is possible to satisfy the slow-roll approximation. Upon expanding perturbatively around the background, the contributions of classical thermal fluctuations to the primordial power spectrum can overwhelm the quantum fluctuations that are usually held responsible for the inhomogeneities of our universe.

\subsection{Power Spectrum}\label{subsec:PS}

In this section we summarise how the thermal fluctuations turn into the super-horizon inflaton fluctuations we observe today. 
A derivation of the results of this section can be found in App.~\ref{App: 2pt}. 
We should point out that  the power spectrum of warm inflation has been calculated many times before, see Refs.~\cite{Graham:2009bf,Bastero-Gil:2011rva,Bastero-Gil:2014raa,Mirbabayi:2022cbt,Montefalcone:2023pvh}, but many of the papers disagree with one another. Our results agree with Ref.~\cite{Mirbabayi:2022cbt}.

During warm inflation, the finite temperature bath has classical thermal fluctuations that eventually imprint themselves on the inflaton. 
We expand the field about the slow-roll background
\begin{equation}
    \phi(\vec{x},t) = \phi(t) + \delta\phi(\vec{x},t).
\end{equation}
The primordial power spectrum is given by the two-point function of the gauge-invariant scalar curvature perturbation
\bea
\zeta &\equiv& -H\frac{\delta\phi}{\dot{\phi}},\\
\langle \zeta_{\vec k}\zeta_{\vec k'}\rangle &\equiv& (2\pi)^3\delta^3(\vec k + \vec k')P(k).
\eea
The equations for the perturbations during warm inflation are derived and numerically solved in App.~\ref{App: 2pt}.  
Roughly speaking, we solve the equation of motion of $\phi$ and the conservation of the stress energy tensor while properly taking into account the thermal fluctuations.
The result using Eq.~\ref{Eq: fluc} and Eq.~\ref{Eq: appfluc} is
\be
P(k) = \frac{1}{k^3}\frac{H^2}{\dot{\phi}^2} \frac{2 c_1 T^4}{f^2}(16\alpha_1^2 + 25 \alpha_2^2)F_2(Q),
\ee
where the function $F_2(Q)$ is shown in Fig.~\ref{Fig: F2} and a numerical approximation to it is given in Eq.~\ref{Eq: F2}. 

\subsection{Thermal Potential}

The coupling of the inflaton to the thermal bath induces both a vacuum and finite-temperature potential.  The vacuum piece can typically be canceled by symmetries such as supersymmetry, while the finite-temperature piece is typically irreducible.  As such, we take the standard warm inflation approach of assuming the vacuum piece has been canceled and only consider the finite-temperature contribution.

%A thermal potential for the inflaton is induced at finite temperature. We must make sure the potential is not so large as to disrupt slow-roll. 
To calculate the thermal potential, we expand the field
\begin{equation}
    \phi = \varphi_{\rm cl} + \delta\phi,
\end{equation}
in Eq.~\ref{Eq: V}. The Lagrangian with this substitution is given by
\bea
    &&\mathcal{L} = |\partial_\mu\chi|^2 - m_\chi^2|\chi^2| - \frac{a(\varphi_{\rm cl})}{4!}\chi^4 + \cdots\\
    &&a(\varphi_{\rm cl}) \equiv \alpha_1 e^{4i\varphi_{\rm cl}/f} + \alpha_2 e^{5i\varphi_{\rm cl}/f}.
\eea
We have kept only the terms relevant for the thermal potential calculation. 
At leading order, we won't need any $\delta\phi$ propagators. 

The potential $V(\varphi_{\rm cl})$ can be calculated from the effective action
\begin{equation}
    \Gamma[\varphi_{\rm cl},\tilde{\varphi}_{\rm cl}] = - \int d^4x \left(V(\varphi_{\rm cl})-\tilde{V}(\tilde{\varphi}_{\rm cl})+\cdots\right),
\end{equation}
which is the generator of one-particle irreducible  diagrams. For completeness, we have included the thermofield double sector, reviewed in App.~\ref{App: thermal}. The potential is given by the vacuum bubble diagrams, and at leading order, one can drop the tilde sector entirely. The leading-order vacuum bubble that gives a potential for $\phi$ is the 3-loop diagram shown in Fig.~\ref{fig:vacbub} :
\begin{figure}[t]
\centering
\includegraphics[width=.4\linewidth]{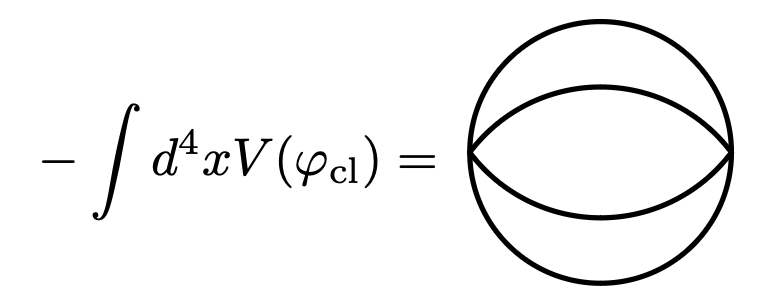}
\caption{The three-loop vacuum bubble which gives the leading-order contribution to the effective potential.}
\label{fig:vacbub}
\end{figure}
\be
V(\varphi_{\rm cl}) = \frac{ |a|^2}{4!}\int d^4y [D_T(x-y)]^4+h.c,
\ee
where $D_T(x-y)$ is the thermal Feynman propagator, the Fourier transform of which is given in Eq.~\ref{eq:FeynProp}. In the limit $m,\muD \rightarrow 0$, this 3-loop diagram has been calculated with dimensional regularization in the imaginary time formalism~\cite{Arnold:1994ps,Braaten:1995jr}, and gives
\begin{equation} \label{Eq: finite T}
    V_T(\varphi_{\rm cl}) = \frac{1.72}{4!}\frac{\alpha_1\alpha_2}{16\pi^2}T^4 \cos(\varphi_{\rm cl}/f),
\end{equation}
where we have chosen to renormalise the couplings at the scale $4\pi T$ in the $\overline{\rm MS}$ scheme.

\subsubsection*{The Effect of the Thermal Potential: }
Finite temperature potentials can have a variety of effects.  
For the more traditional scalar couplings these finite temperature potentials make slow roll more difficult.  
For example, the coupling $y \phi \overline \psi \psi$ would generate a temperature-dependent contribution to the inflaton potential $\delta V \sim y^2 \phi^2 T^2$ at high temperatures.  
This mass can break the $\eta$ slow-roll condition $V'' \ll H^2 (1 + Q)$, making inflation more difficult to achieve.

For periodic scalars, such as the one we are considering, it is a bit more complicated.  To understand this, we first note that the distance the inflaton travels per Hubble patch is typically much larger than $f$, $\dot \phi/fH \gg 1$.  This means that one needs to consider the entire potential shown in Eq.~\ref{Eq: finite T}, as opposed to a perturbative expansion of it.

The effect of the periodic finite-temperature potential is to produce small oscillations on top of the slow-roll potential.  
In other words, instead of changing the slope or curvature of the potential and thus affecting the slow-roll parameters, the inflaton should instead be imagined as rolling down a slope with wiggles on top of it.  
Much like a car rolling down a bumpy road, as long as the kinetic energy of the inflaton is such that it can easily roll over the bumps, the behavior of the background solution will be unaffected~\cite{Hook:2016mqo}.  Namely, as long as the condition
\be
\dot{\phi}^2 \gg \frac{1.72}{4!}\frac{\alpha_1\alpha_2}{16\pi^2}T^4\label{eq:ThermInequality}
\ee
is satisfied, slow roll will proceed as usual.  We have checked this numerically and found that as long as the inequality in Eq.~\ref{eq:ThermInequality} is satisfied by about $\sim 10$, that the usual slow-roll solution is an attractive solution stable under perturbations. More colloquially, even though at any given point $| V''_T | \gtrsim H^2 (1 + Q)$, the rapidly oscillating nature of $V''_T$ means that the effects of $V''_T$ average to zero.  

Despite averaging to zero for the homogeneous piece, there exist instabilities at finite momentum~\cite{Fonseca:2019ypl} when rolling over a bumpy potential. Requiring these instabilities do not lead to a large back-reaction on the slow-roll background provides a further constraint which seems to limit us to the weak regime of warm inflation
\be
\label{eq:PR}
\left(\frac{1.72}{4!}\frac{\alpha_1\alpha_2}{16\pi^2}\frac{T^4}{f^2}\right)^2\left(\frac{\dot{\phi}}{f}\right)^{-3}\lesssim H.
\ee
A derivation of this condition is given in appendix App.~\ref{App: PR}

\subsection{The Data Point}

In this section we present a data point of $\phi^4$ warm inflation where the thermal bath provides the source of fluctuations.
As our model was selected for calculational simplicity, we will not map out all of parameter space and instead simply provide a single data point to show that warm inflation can be realized in this model satisfying all current constraints.

Our example is fairly simple.  
It is our previous model shown in Eq.~\ref{Eq: V} combined with the inflaton potential
\bea
V_{\rm inf} = \frac{\lambda}{4} \phi^4\label{eq:therm1}.
\eea
In order to match the observed cosmological data, one must match the scalar tilt, the scalar power spectrum, and the bound on the scalar to tensor ratio $r$.

Since perturbations are due to the thermal bath, matching the observed scalar power requires (see Eq.~\ref{Eq: fluc}, App.~\ref{App: 2pt}, and Ref.~\cite{Planck:2018jri}) :
\bea
\Delta_R = 2.09\times 10^{-9} = \frac{ c_1 T^4}{\pi^2f^2} \left ( 16 \alpha_1^2 + 25 \alpha_2^2 \right ) \frac{H^2}{\dot \phi^2} F_2(Q).\label{eq:thermalflucs}
\eea
In addition to this, we must match the spectral tilt \cite{Planck:2018jri}
\bea
n_s - 1 &=& - 0.0335 = \frac{d \log \Delta_R}{d N} \nn\\
&=&4 \frac{2-Q}{1+7 Q}\epsilon-6\frac{1+Q}{1+7Q}\eta + \frac{d\log F_2}{d\log Q}\frac{1+Q}{1+7Q}(10\epsilon - 6\eta).\label{eq:tilt1}
\eea
A derivation of this is given in App.~\ref{App: 2pt}. 

Finally, there are a few other constraints we must satisfy. 
\begin{itemize}
    \item The validity of Eq.~\ref{eq:thermalflucs} requires that the thermal fluctuations from warm inflation dominate over the quantum fluctuations:
    \be
    \frac{H^4}{\dot{\phi}^2}\ll\frac{4 c_1 T^4}{f^2} \left ( 16 \alpha_1^2 + 25 \alpha_2^2 \right ) \frac{H^2}{\dot \phi^2} F_2(Q).
    \ee
    \item We consistently made the approximation \be \dot{\phi}/f \ll T,\ee which must be satisfied by our data point. Satisfying this inequality pushes us to low scale inflation. Note that in addition to this, we must make sure $\muD \sim O(\dot{\phi}/f)<m<T.$ 
    \item Ensuring the thermal backreaction is subdominant to the slow-roll potential leads to the inequality in Eq.~\ref{eq:ThermInequality}, and avoiding parametric resonance for the non-zero momentum modes gives the constraint of Eq.~\ref{eq:PR}.
    \item Validity of the effective field theory requires the non-zero temperature not excite mdoes that are out of EFT control
    \be 4 \pi f > 3 T.\ee
    Finally, for perturbation theory in $\alpha_i$ to be valid we require \be \alpha_i < 16\pi^2. \ee
\end{itemize}
All the constraints listed above can be satisfied with the following choices:
\bea
\frac{\alpha_1}{16\pi^2} = 0.443,\qquad \frac{\alpha_2}{16\pi^2}=0.450,\qquad \lambda = 3.98\times 10^{-18}, \quad f = 9.56\times 10^{-4} M_{\rm pl}.
\eea
We also picked $m_\chi/T = 0.55$. These parameters realize warm inflation with
\bea 
\phi = 17.07 M_{\rm pl},\quad Q = 0.027,\quad H = 1.68 \times 10^{-7} M_{\rm pl},\quad T = 8.21 \times 10^{-5}M_{\rm pl}.
\eea
Finally, the prediction for the tensor perturbations in warm inflation is unmodified from cold inflation, so the tensor to scalar ratio is suppressed
\begin{equation}
    r = \left(\frac{\dot{\phi}}{M_{\rm pl}}\right)^2\frac{2f^2 }{ c_1(16\alpha_1^2 + 25\alpha_2^2)T^4F_2(Q)} = 1.36 \times 10^{-6},
\end{equation}
which is well below current bounds. 

We should also mention that this choice of parameters can only support around $40$ e-folds of inflation
$N = \frac{\sqrt{1+Q}}{M_{\rm pl}}\int\limits_{\phi_{\rm end}}^{\phi}\frac{d\phi}{\sqrt{2\epsilon}} \approx 37$.
The number of e-folds required to solve the horizon problem depends both on the Hubble scale during inflation, and the cosmic history of the universe.  Because the Hubble scale for our data point is many orders of magnitude lower than what is typically considered, a lower number of e-folds is required.

%However, by inspection of equations \eqref{}, \eqref{}, and \eqref{}, one can see that by picking \alpha_1 \ll \alpha_2, or vice versa the thermal backreaction can be suppressed 

%In particular, there are thermal corrections to the mass of the inflaton that scale as $\delta m_\phi^2 \propto \alpha_1 \alpha_2 T^4/f^2$, where the presence of both $\alpha_{1,2}$ is due to the phenomenon of collective symmetry breaking.  Slow roll requires $m_\phi \lesssim H$.  As a result, the friction (Eq.~\ref{Eq: boltz fric}) obeys $\gamma \lesssim \delta m^2/T \lesssim H^2/T < H$.  

\section{Conclusions and Further Work} \label{Sec: conclusion}

In this article we have shown that warm inflation utilizing pseudo-scalar inflaton interactions develops a chemical potential. 
In the context of a simple model, we then explained how to calculate both the friction and thermal fluctuations.
We saw that both of these quantities can be drastically altered by the presence of a chemical potential. 

We then discussed models in which the inflaton has generic pseudo-scalar couplings.
We demonstrated that these models develop chemical potentials and estimated the friction terms.
There is still a lot that can be done with these interactions.  
For some interactions, such as the $\phi G \tilde G$ coupling, additional work is required to properly calculate the thermal friction and fluctuations.  For other interactions, such as the $\phi F \tilde F$ coupling with light fermions, the equilibrium configuration has energy injection whose form is not as simple as $\gamma \dot \phi$.
It would be interesting if a warm inflation model could be made where the thermal bath is the Standard Model. 
It should be noted that while warm inflation with the Standard Model has been studied in Ref.~\cite{Berghaus:2025dqi}, it is unclear whether their predictions would survive taking into account the gauge boson chemical potential.

Finally, it would be interesting to explore signals of the thermal classical nature of the fluctuations that led to structure formation.  
Interesting avenues might be the non-Gaussian predictions of warm inflation~\cite{Moss:2007cv,Moss:2011qc,Bastero-Gil:2014raa,Mirbabayi:2022cbt} or its gravitational wave predictions~\cite{Qiu:2021ytc,Klose:2022knn}.  
Additionally, it would be interesting to explore if warm inflation can successfully address some of the typical criticisms of cold inflation without introducing problems of its own.
\section*{Acknowledgments}
We would like to thank Mehrdad Mirbabayi, and Kim Berghaus for helpful discussions.
The authors were supported by NSF grant PHY-2210361 and the Maryland Center for Fundamental Physics.

\appendix

\section{A Lightning Fast Review of the Thermal Field Double} \label{App: thermal}

In this Appendix we give a brief review of how to use the thermal field double to calculate thermal expectation values. 
For a more comprehensive review consult \cite{Khanna:2009zz,Das:1997gg}.
The expectation value of an operator in a thermal state is given by
\bea
\lb A \rb = {\rm Tr} \left ( A \rho \right ) = \frac{1}{Z} \sum_n e^{-\beta E_n} \langle n | A | n \rangle,
\eea
where we have expanded the Hilbert space, $\mathcal{H}$, in the energy eigenbasis, and $Z$ is the partition function. 
The main difficulty in generalizing the techniques of perturbative Quantum Field Theory to finite temperature is that the thermal state is not a pure state; its density matrix is diagonal. 
The thermal field double is a trick that allows one to calculate thermal expectation values like any other quantity in QFT.

The essence of the thermal field double is to have two copies of the theory in question. 
We will suggestively label them the system, with states $n$, and the environment, with states $\tilde n$. 
As long as one only asks questions about the system, the environment is traced over giving the thermal mixed state.  
More precisely we define the ``thermal vacuum"  
\bea
| 0 \rb = \frac{1}{\sqrt{Z}} \sum_n e^{-\beta E_n/2} | n , \tilde n \rangle .
\eea
The expectation value, in this state, of any operator that acts only on the system obeys
\bea
 _\beta \langle 0 | A | 0 \rb &=& \frac{1}{Z} \sum_{n,\tilde n, m, \tilde m}  e^{-\beta (E_n + E_m)/2}  \langle m | A | n \rangle \langle \tilde m | \tilde n \rangle, \\&=& \frac{1}{Z} \sum_n e^{-\beta E_n} \langle n | A | n \rangle = \lb A \rb,
\eea
so we have succeeded in writing thermal expectation values as the expectation value in a pure state.

While not necessary, it is convenient to choose the thermal vacuum, $| 0 \rb $ ,  such that it doesn't evolve in time and has zero energy.
This amounts to saying that the total Hamiltonian $\hat H$ and Lagrangian $\hat \Lag$ obey~\cite{Takahashi:1996zn}
\bea
\hat H = H - \tilde H \qquad \hat \Lag = \Lag - \tilde \Lag.
\eea
If the Lagrangian of the bath is $\Lag(\phi,\phi^\dagger)$, then one can obtain the Lagrangian of the environment using $\tilde \Lag = \Lag(\tilde \phi^\dagger,\tilde \phi)^\dagger$.

We can apply this formalism to the simple harmonic oscillator. 
There are creation and annihilation operators for the system and the environment. 
The thermal vacuum can be obtained by performing a unitary transformation on the ground state of the combined system
\bea
| 0 \rb &=& U(\theta)|0,\tilde{0}\rangle, \\
&=& e^{- \theta_\beta (\tilde a a - \tilde a^\dagger a^\dagger) } | 0 , \tilde{0} \rangle .
\eea
The angle $\theta_\beta$ is given by
\bea
\cosh \theta_\beta &=& \frac{1}{\sqrt{1-e^{-\beta \omega}}}, \qquad \\\sinh \theta_\beta &=& \frac{e^{-\beta \omega/2}}{\sqrt{1-e^{-\beta \omega}}},
\eea
where $\omega$ is the oscillator frequency.

Since the thermal ground state is a unitary transformation acting on the vacuum, the new creation and annihilation operators are Bogoliubov transformations of the familiar creation and annihilation operators with
\bea \label{Eq: Bogoliobov}
a_\beta &=& \cosh \theta_\beta a - \sinh \theta_\beta \tilde a ^\dagger, \\
a_\beta^\dagger &=& \cosh \theta_\beta a^\dagger - \sinh \theta_\beta \tilde a, \\
\tilde a_\beta &=& \cosh \theta_\beta \tilde a - \sinh \theta_\beta a ^\dagger, \\
\tilde a_\beta^\dagger &=& \cosh \theta_\beta \tilde a^\dagger - \sinh \theta_\beta a.
\eea
The thermal creation and annihilation operators obey $a_\beta | 0 \rb = \tilde a_\beta | 0 \rb = 0$.
These expressions have an intuitive understanding.  
Since $a_\beta$ annihilates the vacuum, destroying a particle in the system is equivalent to introducing a particle to the environment. 
In other words, any non-conservation of energy, momentum, and/or particles is just the system exchanging these quantities with the environment. 
The extra factors of $\cosh \theta_\beta$ and $\sinh \theta_\beta$ are merely the normalization factors associated with states that contain several identical particles.

One can calculate the expectation value of the particle number operator in the thermal state
\bea
\,_\beta\lb 0 |a^\dagger a |0\rb &=& \sinh^2\theta_\beta \,_\beta\lb 0| a^\dagger_\beta a_\beta | 0 \rb = \frac{1}{e^{\beta\omega}-1},
\eea
which is the usual Bose-Einstein distribution function. 
It's clear that adding a chemical potential will simply amount to replacing $\omega \rightarrow \omega - \muD$.

The generalisation to field theory is as simple as at zero temperature. 
We know how to decompose any field in terms of the typical creation and annihilation operators, $a$ and $a^\dagger$.  
We can use Eq.~\ref{Eq: Bogoliobov} and its inverse to write these fields in terms of the thermal creation and annihilation operators acting on the thermal bath, $a_\beta$ and $a_\beta^\dagger$.  
For our application we should include a chemical potential, $\muD$, for the particle and, $-\muD$, for the antiparticle.
More explicitly,
\bea
\chi(x) &=& \int \frac{d^3 p}{(2 \pi)^3 \sqrt{2 \omega_p}} \left ( a_p e^{-i p x} + b^\dagger_p e^{i p x} \right ) , \\
 &=& \int \frac{d^3 p}{(2 \pi)^3 \sqrt{2 \omega_p}} \left ( (u_\beta(\muD) a_{p,\beta} + v_\beta(\muD) \tilde a^\dagger_{p,\beta}) e^{-i p x} + (u_\beta(-\muD) b^\dagger_{p,\beta} + v_\beta(-\muD) \tilde b_{p,\beta}) e^{i p x} \right ) \nn \\
u_\beta(\muD) &=& \frac{1}{\sqrt{1-e^{-\beta (\omega_p - \muD)}}} \qquad v_\beta(\muD) = \frac{e^{-\beta (\omega_p - \muD)/2}}{\sqrt{1-e^{-\beta (\omega_p - \muD)}}},
\eea
where we have left some of the $p$ dependence implicit.
At this point, the calculation proceeds as any other standard Quantum Field Theory calculation. 

For applications in the main text we will require various expressions for the thermal field propagator. 
Firstly, the usual Feynman propagator will be used to calculate the thermal back-reaction on the slow-roll potential
\bea
    iD_F(p,T) &\equiv& \int d^4x e^{ip\cdot(x-y)} \,_\beta\lb 0 | T\{\chi(x)\chi^\dagger(y)\}|0\rb\\
    &=& \frac{i}{p^2-m^2+i\epsilon} + (2\pi) \delta(p^2-m^2)\left(\theta(p^0)\bar{f}_p+\theta(-p^0)f_p\right)\label{eq:FeynProp},
\eea
where the distribution functions are defined as in Eqns.~\ref{eq:disfuncs1} and \ref{eq:disfuncs2}. It's interesting to note that the effect of the non-zero temperature is to simply modify the propagator by the addition of a single term that depends on the distribution function. 

We use the in-in formalism to calculate the friction term in the equation of motion. 
This requires the use of propagators without time-ordering. 
Because unitarity is simpler in the time domain~\cite{tHooft:1979hnm}, it is convenient not to Fourier transform in time giving
\bea
    \,_\beta\lb 0 |\chi^\dagger(x)\chi(y)|0\rb &=& \int \frac{d^3p}{(2\pi)^3 2\omega_p}\left((1+\bar{f}_p)e^{-i p\cdot(x-y)}+f_p e^{ip\cdot(x-y)}\right)\label{eq:Prop1},\\
    \,_\beta\lb 0 |\chi(x)\chi^\dagger(y)|0\rb &=& \int \frac{d^3p}{(2\pi)^3 2\omega_p}\left((1+f_p)e^{-i p\cdot(x-y)}+\bar{f}_p e^{ip\cdot(x-y)}\right)\label{eq:Prop2}.
\eea
One can easily pick out initial and final states in the Boltzmann equation from this form of the propagator.

\section{A lightning Fast Review of the In-in Formalism} \label{App: in-in}

In this Appendix we give a lightning fast review of the in-in formalism.  We will be a bit sloppy with the $i \epsilon$ prescription as it will not matter for the calculations we do in this paper.  
When one first learns Quantum Field Theory, we are taught in-out expectation values defined as 
\bea
\langle 0 | T \left( e^{- i \int^{\infty}_t H^{\rm int}(t_2) dt_2} \right )  O(t) T \left( e^{- i \int_{-\infty}^t H^{\rm int}(t_1) dt_1} \right ) | 0 \rangle .
\eea
The ket is the vacuum in the past, time evolved to when the measurement occurs.  This is called the in state.  Meanwhile, the bra is the vacuum in the future, time evolved backwards to when the measurement occurs.  This is called the out state.  While completely un-intuitive, this object is easy to calculate and related to physical processes such as cross sections.

Even further back, when one first learns quantum mechanics, expectations values are instead evaluated as
\bea
\langle \Psi(t) | O(t) | \Psi(t) \rangle &\sim& \langle 0 | \left ( T \left( e^{- i \int_{-\infty}^t H^{\rm int}(t_2) dt_2} \right ) \right )^\dagger  O(t) T \left( e^{- i \int_{-\infty}^t H^{\rm int}(t_1) dt_1} \right ) | 0 \rangle \nn \\
&=& \langle 0 |  \overline  T \left( e^{i \int_{-\infty}^t H^{\rm int}(t_2) dt_2} \right )  O(t) T \left( e^{- i \int_{-\infty}^t H^{\rm int}(t_1) dt_1} \right ) | 0 \rangle , \label{Eq: in-in}
\eea
where $\overline T$ represents anti-time ordering.  Because the bra and ket state are both obtained by time-evolution of a state from the past to the present, this is called an in-in expectation value.

The in state and the out state are related by time evolution of the vacuum from $t=-\infty$ to $\infty$.  As such, in-in and in-out calculations are often just related to each other by a phase.  For time dependent vacua and Lagrangians, the situation can become significantly more complicated.

A particularly useful way in which to write Eq.~\ref{Eq: in-in} is~\cite{Weinberg:2005vy,Baumgart:2019clc}
\bea
\langle O(t) \rangle = \sum_{V=0}^\infty (-i)^V \int_{t_0}^t dt_v \cdots \int_{t_0}^{t_3} dt_2 \int_{t_0}^{t_2} dt_1 \langle [[ \cdots [ \mathcal{O},H_{\rm int}(t_v)] \cdots ,H_{\rm int}(t_2)] , H_{\rm int}(t_1)] \rangle. \,\,\,\,\label{Eq: nest}
\eea
It turns out that this way of writing expectation values in terms of commutators is particularly conducive to calculations using the thermal field double. Both in-in and in-out calculations give exactly the same result, because in the thermal field double calculation the thermal vacuum was chosen to be time invariant with energy equal to zero. 
However, in-in calculations are computationally easier. 
The reason for this is that as shown in Eq.~\ref{Eq: nest}, in-in calculations involve retarded propagators while in-out calculations only ever use Feynman propagators.  
%This can easily be seen by expanding Eq.~\ref{Eq: in-in} to leading order in perturbations.
Because the bath and environment fields commute
\bea
[ \phi , \tilde \phi ] = 0,
\eea
when doing in-in calculations, the environment simply does not show up.  This renders calculations much easier to do in the in-in formalism as opposed to the in-out formalism.  Of course, the two approaches give the same answer.

\section{Parametric Resonance} \label{App: PR}

In this appendix we give a quick derivation of the condition Eq.~\ref{eq:PR}. 
For a more detailed discussion consult \cite{Fonseca:2019ypl}, where the estimates given here are checked numerically. 
In the presence of the thermal effective potential Eq.~\ref{Eq: finite T} the equation of motion for the fluctuation is given by 
\be
\delta\ddot{\phi}_k+\Gamma\delta\dot{\phi}_k+\left(\frac{k^2}{a_0^2}e^{-2Ht}-A\cos(\omega t)\right)\delta\phi_k = 0,
\ee
where we have dropped the contribution to the fluctuations from interactions with the radiation sector to zoom in on the effect of parametric resonance. 
The various constants are given by
\be\Gamma = 3H + \gamma;\quad A=\frac{1.72}{4!}\frac{\alpha_1\alpha_2}{16\pi^2}\frac{T^4}{f^2}; \quad\omega = \dot{\phi}/f.\ee
This equation has an instability when
\be \frac{k}{a_0}e^{-Ht_*}\approx \frac{\omega}{2},\ee
because the final term will have a component that acts like a forcing term on resonance. 
We have defined $t_*$ as the time at which this particular k-mode is on resonance. 

We assume that the length of time a given k mode is unstable is short relative to Hubble, $t=t_*+\delta t$, and that the friction term can be neglected during the instability \be \frac{k^2}{a_0^2}e^{-2Ht} \approx \frac{\omega^2}{4}(1-2H\delta t);\qquad \frac{\Gamma}{\omega}\ll \frac{A}{\omega^2}.\ee
During the instability, the equation of motion for the fluctuations resembles the Matthieu equation, and the unstable mode grows exponentially \cite{Fonseca:2019ypl} \be \delta\phi_k(t)\sim \delta\phi_k(t_*)\exp\left(\frac{A}{2\omega}(t-t_*)\right).\ee
This equation is true exactly on resonance, and therefore gives an upper bound on the exact growth rate.

To estimate the size of the enhancement we need to know how long each mode is unstable. 
It turns out that each mode is unstable as long as \be \frac{\omega^2}{4} - \frac{A}{2} < \frac{\omega^2}{4}(1-2H\delta t) < \frac{\omega^2}{4} + \frac{A}{2}.\ee
We use this condition to estimate the length of time the mode is unstable as \be\Delta t = \frac{1}{H}\frac{2A}{\omega^2}.\ee
Note here that $A\ll \omega^2$ is the condition Eq.~\ref{eq:ThermInequality} that the inflation has enough kinetic energy to roll over the thermal potential, so this confirms our assumption that the instability is short relative to Hubble. 
If we plug this in we find that during the instability the non-zero k mode grows by a factor \be \exp\left(\frac{1}{H}\frac{A^2}{\omega^3}\right) \implies \frac{A^2}{\omega^3} \ll H, \ee
which is the inequality given in Eq.~\ref{eq:PR}\footnote{A slightly more sophisticated estimate, taking into account that the mode is slightly off resonance during the instability, gives a factor of $\pi/4$ as opposed to the factor of 1, and fits the numerics remarkably well \cite{Fonseca:2019ypl}.}.

\section{The Fluctuation Dissipation Theorem} \label{App: fluc}

In this appendix we present the derivation of the fluctuation dissipation theorem and show that our results obey it.  While the derivation of the fluctuation dissipation theorem with a chemical potential for a conserved charge is well known, we are interested in the theorem in the presence of a chemical potential for a non-conserved charge.  At leading order in the symmetry breaking coupling, there is no difference between the two situations.  As this is all we need to show the validity of our results, we do not push further.  It would be interesting to know what is the full form of the fluctuation dissipation theorem in the presence of chemical potentials for non-conserved charges.

\subsection{Derivation of the Theorem}

Let us first present a quick derivation of the fluctuation-dissipation theorem with a chemical potential. The presentation will follow the style of \cite{Laine:2016hma}. The thermal expectation value of an operator is given by
\begin{equation}
    \langle \mathcal{O}(x)\rangle \equiv \frac{1}{Z}\textrm{Tr}\left[\mathcal{O}(x)e^{-\beta(H-\muD Q)}\right],
\end{equation}
where $Q$ is the charge operator, which in our model is the $\chi$ number operator. For an operator with a definite charge under $Q$
\begin{equation}
    [\mathcal{O},Q] = q\mathcal{O}.\label{eq:commrel}
\end{equation}
The theorem relates the following 3 thermal expectation values
\bea
    \Pi_\mathcal{O}^>(k) &\equiv& \int d^4x e^{ik\cdot x}\langle \mathcal{O}(x)\mathcal{O}^\dagger(0)\rangle,\\
    \Pi_\mathcal{O}^<(k) &\equiv& \int d^4x e^{ik\cdot x}\langle \mathcal{O}^\dagger(0)\mathcal{O}(x))\rangle,\\
    \rho_\mathcal{O}(k) &\equiv& \int d^4x e^{ik\cdot x}\frac{1}{2}\langle [\mathcal{O}(x),\mathcal{O}^\dagger(0)]\rangle.
\eea
For the purpose of this derivation we must assume $[H,Q]=0$. Although this isn't true because the couplings break $\chi$ number conservation, it is true in the free theory, and can be used at leading order in perturbative calculations. We can insert $1 = \sum_n | n \rangle \langle n |$ twice into the first two expectation values to obtain
\bea
\Pi^>_{\mathcal{O}}(k) &=& \frac{1}{Z}\sum\limits_{m,n}\int d^3x e^{-i\vec{k}\cdot\vec x}(2\pi)\delta(k^0+E_m-E_n)e^{-\beta E_m}\langle m| e^{\beta\muD Q}\mathcal{O}(\vec{x})|n\rangle\langle n|\mathcal{O}^\dagger(0)|m\rangle,\,\,\,\,\,\,\,\\
\Pi^<_{\mathcal{O}}(k) &=& \frac{1}{Z}\sum\limits_{m,n}\int d^3x e^{-i\vec{k}\cdot \vec x}(2\pi)\delta(k^0+E_m-E_n)e^{-\beta E_n}\langle m| \mathcal{O}(\vec{x})e^{\beta\muD Q}|n\rangle\langle n|\mathcal{O}^\dagger(0)|m\rangle,\,\,\,\,\,\,\,
\eea
where we used $\mathcal{O}(\vec{x},t)= e^{iHt}\mathcal{O}(\vec{x})e^{-iHt}$ in the intermediate steps. At this point one can use the commutation relation shown in Eq.~\ref{eq:commrel} to find $\mathcal{O}(\vec{x})e^{\beta\muD Q} = e^{\beta\muD(Q+q)}\mathcal{O}(\vec{x})$, giving a relationship between the first two expectation values. Combining this with the fact that the third expectation value is half of the difference of the first two we obtain the fluctuation-dissipation theorem
\be
     \rho_\mathcal{O}(k) = \frac{1}{2}\left(e^{\beta(k^0-\muD q)}-1\right)\Pi^<_\mathcal{O}(k).\label{eq:FlucDissThm}
\ee

\subsection{Consistency of our Calculation}

We can check if the calculations of Sec.~\ref{Sec: toy} are consistent with this theorem. Focus on the first term of Eq.~\ref{Eq: 1 point}
\bea
-\gamma\dot{\phi}&\supset&-\frac{4}{f}\frac{\alpha_1^2}{(4!)^2}\int d^3y\int_{-\infty}^{t_x}dt_y e^{4i\frac{\dot{\phi}}{f}(t_x-t_y)}\langle [\chi^4(x),\chi^{\dagger 4}(y)]\rangle\\
&=& \frac{4}{f}\frac{\alpha_1^2}{(4!)^2}\int d^4y e^{-4i\frac{\dot{\phi}}{f}(t_y-t_x)}\frac{1}{2}\langle [\chi^{\dagger 4}(y),\chi^4(x)]\rangle\\
&=& \frac{4}{f}\frac{\alpha_1^2}{(4!)^2}\rho_{\chi^\dagger}(-k_1),
\eea
where $k_1 = (4\dot{\phi}/f,\vec 0)$, and we have extended the integral to infinity by recalling from the main text that after adding the hermitian conjugate the integral over time gave an energy conservation delta function. If we add in all the other contributions to the friction term we find
\begin{equation}
    -\gamma\dot{\phi} = - \frac{4}{f}\frac{\alpha_1^2}{(4!)^2}\left(\rho_\chi(k_1)-\rho_{\chi^\dagger}(-k_1)\right) - \frac{5}{f}\frac{\alpha_2^2}{(4!)^2}\left(\rho_\chi(k_2)-\rho_{\chi^\dagger}(-k_2)\right),\label{eq:Fricapp}
\end{equation}
where $k_2 = (5\dot{\phi}/f,\vec 0)$. 

We can read off from Eq.~\ref{Eq: A} that the fluctuations
\bea
A = \frac{16\alpha_1^2}{(4!)^2f^2}\left(\Pi_\chi^<(k_1)+\Pi^<_{\chi^\dagger}(-k_1)\right) +\frac{25\alpha_2^2}{(4!)^2f^2}\left(\Pi_\chi^<(k_2)+\Pi^<_{\chi^\dagger}(-k_2)\right).\label{eq:Flucapp}
\eea
Now consider the fluctuation dissipation theorem (Eq.~\ref{eq:FlucDissThm}) at leading order by noting that $q=\pm 4$
\bea
\rho_\chi(k) &\approx& \frac{1}{2}\beta(k^0-4\muD)\Pi^<_{\chi}(0),\label{eq:fluc1}\\
\rho_{\chi^\dagger}(k) &\approx& \frac{1}{2}\beta(k^0+4\muD)\Pi^<_{\chi^\dagger}(0)\label{eq:fluc2}.
\eea
Note that we have applied the fluctuation dissipation theorem separately to the $\pm 4$ charged operators because these obey Eq.~\ref{eq:commrel} separately but not as a sum.
This is expression is another way to see what we mentioned in the main text that at leading order the chemical potential leaves the fluctuation unaffected while reducing the friction. By setting the argument of $\Pi$ to zero we see that the expression for $A$ is completely independent of $\muD$.

Plugging Eq.~\ref{eq:fluc1} and Eq.~\ref{eq:fluc2} into Eq.~\ref{eq:Fricapp}, and setting $k_1,k_2=0$ in Eq.~\ref{eq:Flucapp} gives exactly the relationship between the friction term and the fluctuations that we found in Eq.~\ref{eq:FlucFinal}. Also, notice that if $\muD$ is set to zero in Eqs.~\ref{eq:fluc1} and \ref{eq:fluc2} then we would obtain $A = 2\gamma T$, which shows that $\muD=0$ was an important implicit assumption in previous uses of the fluctuation-dissipation theorem in warm inflation.

\section{From Thermal Fluctuations to Inflaton Fluctuations} \label{App: 2pt}

In this Appendix, we derive the results used in Sec.~\ref{subsec:PS} of the main text.
This work is necessarily partly numerical as the relevant Greens functions cannot be obtained completely analytically.  
Additionally, as mentioned in the main text, while this aspect of warm inflation has been addressed many times before, see Refs.~\cite{Graham:2009bf,Bastero-Gil:2011rva,Bastero-Gil:2014raa,Mirbabayi:2022cbt}, many of the papers disagree with each other.  
The disagreement may in part be numerical, however many of the papers disagree with each other on what is the form of the equation to solve.~\footnote{All references agree that the equations to use are the conservation of the stress energy tensor, Eq.~\ref{Eq: d4}, and the equation of motion for the inflaton, Eq.~\ref{Eq: eom}.
We disagree with Ref.~\cite{Graham:2009bf,Bastero-Gil:2011rva,Bastero-Gil:2014raa,Montefalcone:2023pvh} because in Eq.~\ref{Eq: d5}, we include the fluctuation term $\xi$ coming from using the inflaton equation of motion, Eq.~\ref{Eq: eom}, whereas they drop this term.  Additionally in Eq.40 of Ref.~\cite{Graham:2009bf}, they copied Eq.A12 of Ref.~\cite{Moss:2007cv} incorrectly dropping the additional term $2 \Gamma \dot \phi \delta \dot \phi$, leading to additional differences with the rest of the references.
 } 
In this Appendix, we discuss our approach to the problem, which follows Ref.~\cite{Mirbabayi:2022cbt}.

In what follows, we use the mostly positive metric
\bea
ds^2 = - dt^2 + a(t)^2 dx^2,
\eea
and use dots to signify time derivatives and $'$ to signify derivatives with respect to the conformal time $\eta = -1/(aH)$.  We will work in spatially flat gauge and in the limit $M_p \rightarrow \infty$.  Additionally, we will take the temperature dependence of the frictional coefficient to be $\gamma \propto T^c$ and the equation of state of the thermal bath to be $w$.

\subsection{The Differential Equations} 

We first discuss the differential equation that turns thermal fluctuations into inflaton fluctuations.  
In the limit where we are considering $k$ modes corresponding to wavelengths much longer than the correlation length of the thermal bath, we can treat the fluid as an ideal fluid.  
As a result there are three scalar fluctuations that we need to consider, the inflaton fluctuations, the temperature fluctuations, and fluctuations in the local velocity of the bath.  
We express these three fluctuations as
\bea \label{Eq: flucs}
\phi = \phi_0 + \dot \phi_0 \Phi \qquad T_{\rm bath,0}^0 = - \rho_0 (1 + \E ) \qquad \partial_i T_{\rm bath,0}^i = - \frac{4 \rho_0}{3 a^2}\nabla^2 \Psi .
\eea

To solve for these three variables, we have the three equations coming from the equation of motion of $\phi$ and the conservation of the stress energy tensor.  
The equation of motion for $\phi$ is
\bea \label{Eq: eom}
D^2 \phi - V,_{\phi}(\phi) = \gamma u^\mu \partial_\mu \phi + \xi ,
\eea
where $u^\mu$ specifies the rest frame of the thermal bath, $\xi $ is the noise term, and $D$ is the covariant derivative. Meanwhile conservation of the stress energy tensor gives
\bea\label{Eq: d4}
D_\mu T^{\mu}_\nu &=& D_\mu T^{\mu}_{\rm bath,\nu} + D_\mu T^{\mu}_{\phi,\nu}   =  0 \\ \label{Eq: d5}
D_\mu T^{\mu}_{\rm bath,\nu}  &=& -  D_\mu T^{\mu}_{\phi,\nu}   =  - \partial_\nu \phi \left ( \gamma u^\mu \partial_\mu \phi + \xi  \right )
\eea
where in the last line we have used the equation of motion, Eq.~\ref{Eq: eom}.  Removing the homogeneous pieces of these equations, we find the equations that govern the evolution of the perturbations to be
\bea \label{Eq: all EOM}
& & \Phi'' - \left ( \frac{2}{\eta} + \frac{\gamma}{H \eta} \right ) \Phi' + k^2 \Phi = -\frac{c \gamma \E}{4 H^2 \eta^2} + \frac{\xi}{H^2 \eta^2 \dot \phi_0}, \\
& & \E' - \frac{3 ( 1 + w)}{\eta} (1 - \frac{c}{4} ) \E + \frac{4 k^2 H \eta \Psi}{3} - 6 H (1 + w) \Phi' = \frac{3 (1+w)}{\gamma \eta \dot \phi_0} \xi \nn, \\
& &\Psi' - \frac{3}{\eta} \Psi - \frac{3 w}{4 H \eta} \E - \frac{9 (1 +w)}{4 \eta} \Phi = 0. \nn
\eea

\subsection{Numerical Results} 
To proceed further, we need to solve Eqs.~\ref{Eq: all EOM} subject to the thermal fluctuations
\bea \label{Eq: thermal fluc}
\langle \xi_k(\eta) \xi_{k'}(\eta') \rangle = A H^4 \eta^4 \delta(\eta - \eta') (2 \pi)^3 \delta^3(k + k'). 
\eea
The observed density fluctuations are written in terms of the gauge invariant perturbation $\zeta$.  After gauge fixing, $\zeta_k = -\frac{H}{\dot \phi_0} \delta \phi_k$,  the observed density fluctuations are
\bea \label{Eq: appfluc}
\langle \zeta_k \zeta_{k'} \rangle 
%= 2.09 \times 10^{-9} \frac{(2 \pi)^3 \delta^3(k + k')}{k^3} 
= \frac{A H^2}{\dot \phi_0^2} F_2(Q) \frac{(2 \pi)^3 \delta^3(k + k')}{k^3},
\eea
where $Q \equiv \gamma/3H$. All that remains for us to do is determine numerically the function
\bea \label{Eq: F2definition}
F_2(Q) = k^3 \int_{-\infty}^0 d\eta\,\eta^4 G_{\Phi_k}(0,\eta)^2 ,
\eea
where $G_{\Phi_k}(\eta,\eta')$ is the Greens function for the inflaton fluctuations $\Phi$ obtained by solving Eqs.~\ref{Eq: all EOM} with the replacement $\xi \rightarrow \dot \phi_0 \delta(\eta - \eta')$.

\begin{figure}[t]
\centering
\includegraphics[width=.48\linewidth]{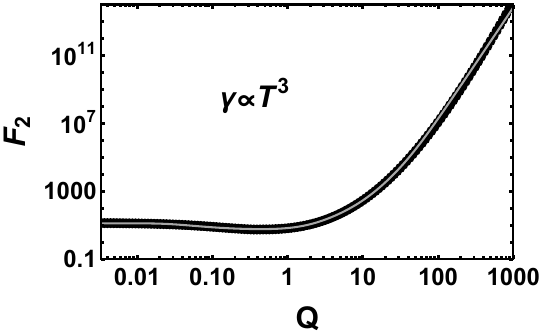}
\includegraphics[width=.48\linewidth]{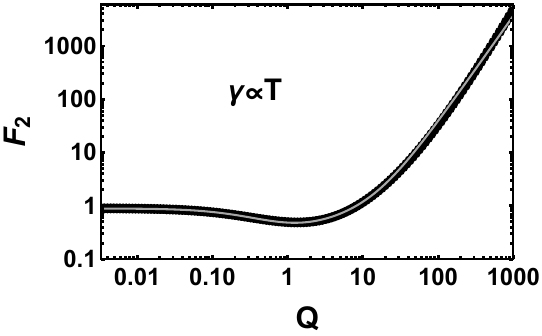}
\includegraphics[width=.48\linewidth]{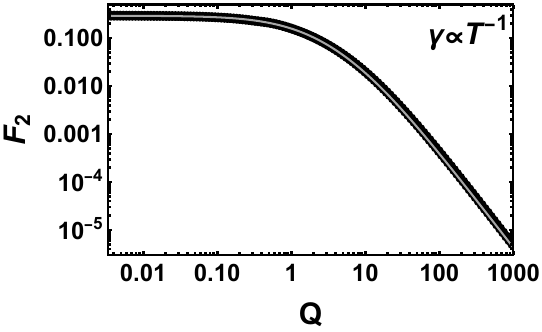}
\caption{A comparison of the numerical value of $F_2$ versus the numerical fits shown in Eq.~\ref{Eq: F2}. The fits are shown in the solid lines while the dots represent the numerical solution.}
\label{Fig: F2}
\end{figure}

Since our approach can generate $\gamma \sim T^c$ with $c =3, 1, -1$, we will be considering these three options when studying the two point function numerically.  In all cases, we will restrict ourselves to a radiation bath obeying $ w = 1/3$.  We find a good approximation for our numerically obtained $F_2$ are
\bea \label{Eq: F2}
F_2(Q) \approx \begin{cases}
    5 + 7.9 \tanh \left ( \frac{1}{30 Q} \right ) + 1.5 Q^2 + 7.8 \times 10^{-3} Q^{4.2} + 1.0 \times 10^{-6} Q^{6.6}, & \gamma \sim T^3\\
    \frac{1}{1.1 + 3.2 Q} + 0.25 \sqrt{Q} + 3.6 \times 10^{-3} Q^2, & \gamma \sim T\\
    \frac{1}{3.4 + 2.7 Q + 0.23 Q^2}, & \gamma \sim T^{-1}
\end{cases}
\eea
We compare these analytic approximations to the numerical solutions of $F_2$ in Fig.~\ref{Fig: F2}. 

We now discuss some of the physics that goes into these solutions.  In particular, we will focus on the behavior of the modes present at large $Q = \gamma/3H$ as can be seen in Fig.~\ref{Fig: F2}.
The delta function source in Eq.~\ref{Eq: thermal fluc} produces fluctuations at all frequencies and the goal is to identify and understand the dynamics of the modes that eventually dominate the inflaton fluctuations we observe today.  We will discuss things chronologically as the physical momentum goes from large to small.

\paragraph{$c > 0$ : } 
When the physical momentum is larger than $\sqrt{\gamma H}$, inflaton perturbations are underdamped oscillators.  In this region of parameter space, the WKB approximation shows that these modes dilute away.  As such inflaton modes that begin their life as thermal fluctuations with large momenta like this are necessarily subdominant.

Next we consider when the physical momentum is smaller than $\sqrt{\gamma H}$ but still larger than Hubble.  In this region, there is an exponential instability.  The critical consideration behind this instability is how does a $\delta \phi$ fluctuation deposit energy and momentum into the fluid.
In the momentum regime we are considering, momentum deposition is more important than energy deposition, and most of energy goes into pressure gradients.  Because pressure points outwards, a large pressure gradient causes the temperature to drop.  As a result, an upward fluctuation in $\delta \phi$ deposits a lot of momentum and causes temperature to decrease, which causes the friction to decrease and thus speeding up the growth of $\delta \phi$ leading to a runaway.

When the physical momentum is smaller than Hubble, Hubble friction is more important than anything else in the story.  As is familiar, Hubble friction brings perturbations to a grinding halt.  As a result, the only superhorizon mode that survives until late times is the one that becomes a constant.

%When the physical momentum is smaller than Hubble, energy is deposition is important than momentum deposition.  In this case there is no instability.  An upward fluctuation in $\delta \phi$ is more energy in the inflaton and causes temperature to increase, which causes the friction to increase and thus slow down the growth of $\delta \phi$, a negative feedback cycle.  As a result, the only superhorizon mode that survives until late times is the one that becomes a constant.

We can now trace the origin of the perturbations that dominate the end result.  These modes were originally produced by the delta function thermal fluctuation when their momentum was of order $\sqrt{\gamma H}$.  These modes grow in size until they are redshifted out of the instability band and become superhorizon.  At that point they freeze out and become constant.

%{\bf If we can't obtain the correct answer, then we need to remove this.}

%We can double check our understanding of the origin of perturbations by analytically deriving the large x behavior of $x^6$ ($x^2$) when $c=3$ ($c=1$).  When $H < k_{\rm physical} < \sqrt{\gamma H}$ and $\gamma \gg H$, the third equation of Eq.~\ref{Eq: all EOM} sets $\E = -12 H\Phi$.  Thus, the Greens functioaqsn of interest is
%\bea
%\eta \gamma H G(\eta,\eta')' + (3 c \gamma H - H^2 \eta^2 k^2) G(\eta,\eta') = \delta ( \eta - \eta') \quad \Rightarrow \quad
%G(0,\eta') \propto \frac{\eta'^{3c-1}}{\gamma} e^{i k \eta'} .
%\eea
%We can now use Eq.~\ref{Eq: F2definition}, to integrate up to when the instability first appears and obtain the scaling
%\bea
%F_2(x) \sim \gamma^{3c-1}
%\eea
%in the large $\gamma$ limit.  This analytic estimate of the scaling agrees well with the numerical results.

\paragraph{$c < 0$ : } 

The $c < 0$ story is similar to the $c > 0$ except for in the instability region, $H < k_{\rm physical} < \sqrt{\gamma H}$.  As before, momentum deposition is the dominant effect and so the logic is similar to before.  An upward fluctuation in $\delta \phi$ deposits a lot of momentum and causes temperature to decrease, which causes the friction to {\it increase} (as opposed to $c>0$ which caused a decrease) and slowing down the growth of $\delta \phi$ leading to a feedback mechanism that stabilizes any perturbation.  Thus there is no exponential growth and only friction.

Because when $c < 0$, there is no enhancement of modes, the relevant modes are dominantly produced when $k_{\rm physical} \sim H$ so as to minimize frictional effects diminishing the amplitude before freeze-out.  In this almost slow roll limit, the $\gamma$ dependence of the Greens function is clear since the only place that $\gamma$ appears is as $\xi/\gamma$.  So that the Greens function $\sim 1/\gamma$ and $F_2 \sim 1/\gamma^2$ as seen in Fig.~\ref{Fig: F2}.

\subsection{Derivation of the Spectral Tilt}

Fitting to experimental data requires calculating the spectral tilt
\be n_s-1 \equiv \frac{d \log \Delta_R}{d N},\ee
where $dN = Hdt$. From Eq.~\ref{Eq: appfluc} we see that the spectral tilt is given by
\be
n_s -1 = 2\frac{d\log H}{dN} - 2 \frac{d \log \dot{\phi}}{d N} + \frac{d\log A}{dN} + \frac{d \log F_2}{d\log Q}\frac{d\log Q}{dN}\label{eq:tiltapp}.
\ee
We must write the derivative here in terms of the slow-roll parameters. By directly taking the derivatives of the LHS in Eqs.~\ref{eq:sr1}, and \ref{eq:sr2} one finds
\bea
&&\frac{d \log H}{dN} = - \epsilon\\
&& \frac{d\log \dot{\phi}}{dN} = \epsilon - \eta - \frac{Q}{1+Q}\frac{d\log Q}{dN}.
\eea
To write the final term in terms of $\epsilon$ and $\eta$ we use $Q \sim T^c/H$ to write
\be \frac{d \log Q}{d N} =\epsilon + c \frac{d\log T}{d N}.\ee
Finally one can recall that $\rho_r \sim T^4$, and use Eq.~\ref{eq:sr3} to find
\be \left(4 + c \frac{Q-1}{Q+1}\right) \frac{d \log T}{d N} = \frac{3+Q}{1+Q} \epsilon - 2 \eta.\ee
This is easily rearranged to give $d\log T/dN$ in terms of $\epsilon$, and $\eta$. If we specialise to $c=3$, and plug into Eq.~\ref{eq:tiltapp}, then we arrive at Eq.~\ref{eq:tilt1}.

\section{Linear Response Theory and the Adiabatic Approximation} \label{App: linear}

Most other approaches to the perturbations and fluctuations of warm inflation utilize linear response theory~\cite{Berera:2002sp,Bastero-Gil:2010dgy}.  In this section we briefly review linear response theory and the adiabatic approximation emphasizing how they are  simply related to how we went about calculating expectation values.

Let us take an interaction between the bath and the inflaton to be $V_{\rm int}(\phi, \chi) = \epsilon f(\phi) g(\chi)$.  In the approach we took in the main text, we would write the equation of motion as
\bea \label{Eq: Vint}
\Box \phi + 3 H \dot \phi + V'(\phi) = - \epsilon_{\rm in}\langle f'(\phi) g(\chi) \rangle_{\rm in,\beta} ,
\eea
and subsequently utilize perturbation theory to calculate the expectation value of the in-in correlator.  Linear response theory is when one calculates the expectation value in Eq.~\ref{Eq: Vint} to first order in perturbation theory.  To first order in perturbation theory with $V_{\rm int}$, the above expectation value is 
\bea
-_{\rm in}\langle f'(\phi) g(\chi) \rangle_{\rm in,\beta} = i \epsilon \int d^4 y  \theta ( t_x - t_y )\lb [  f'(\phi_x) g(\chi_x) , f(\phi_y) g(\chi_y)] \rb + \mathcal{O}(\epsilon^2),
\eea
where the subscripts indicate what fields are functions of $x$ versus $y$ and we are using the subscript $_{\rm in}$ to indicate what expectation values are calculated including the $e^{-i \epsilon V_{\rm int} t}$ and which are not.  We now massage the expression a bit by plugging in the background values of $\phi$ and adding and subtracting $\delta V'(\phi_x)$
\bea
-_{\rm in}\langle f'(\phi) g(\chi) \rangle_{\rm in,\beta} &=& i \epsilon  f'(\phi_x) \int d^4 y  \theta ( t_x - t_y ) (f(\phi_y) - f(\phi_x)) \lb [  g(\chi_x) , g(\chi_y)] \rb - \delta V'(\phi_x) \nn \\
\delta V'(\phi_x) &=& -i \epsilon  f'(\phi_x) \int d^4 y  \theta ( t_x - t_y )  f(\phi_x) \lb [  g(\chi_x) , g(\chi_y)] \rb . \label{Eq: linearresponse}
\eea
The first term in this expression is what is typically used in linear response theory with $\delta V(x)$ being absorbed into the definition of $V(x)$.  We can now summarize the difference between the two approaches, which we note will give exactly the same expression.  Linear response theory is the $\epsilon^1$ term in perturbation theory with ``local" effects (terms that only depend on $\phi_x$) pulled into the potential.  These last terms were extremely briefly noted in Sec.~\ref{Sec: thermalfric}.  The remaining terms are ``non-local", namely they can contribute to derivatives terms of the inflaton, e.g.  $\dot \phi_x$.

Finally, when evaluating Eq.~\ref{Eq: linearresponse}, the adiabatic approximation is typically used.  The adiabatic approximation is
\bea
f(\phi_y) - f(\phi_x) &\approx& f'(\phi_x) \dot \phi_x (t_y - t_x) \nn \\
- \epsilon_{\rm in}\langle f'(\phi) g(\chi) \rangle_{\rm in,\beta} &=&  - \gamma \dot \phi_x \nn \\
\gamma &=& i \epsilon^2  f'(\phi_x)^2 \int d^4 y  \theta ( t_x - t_y ) (t_x - t_y) \lb [  g(\chi_x) , g(\chi_y)] \rb.
\eea
This Taylor series in time is equivalent to our Taylor series $\dot \phi/f \ll T$ that was done in e.g. Eq.~\ref{Eq: dndt}.  Because of the special manner in which our inflaton coupled to the $\chi$ field, this approximation is not needed as the time integrals give nice delta functions without the need of a Taylor series.  However, the expansion makes the integrals easy to do, which is why we did the expansion.

In summary, as long as back reaction is taken into account, what we have done is identical to the typically considered linear response theory with the adiabatic approximation.  We simply framed the calculation differently.

\bibliographystyle{JHEP}
\bibliography{biblio}

\end{document}